\documentclass[10pt]{article}
\usepackage{amsmath}
\usepackage{amsfonts}
\usepackage{amssymb}
\usepackage{graphicx}
\usepackage[table,xcdraw]{xcolor}
\usepackage{orcidlink}
\usepackage{xstring}
\newcommand{\hrefbibentry}[2]{%
  \IfStrEq{#1}{}{#2}{%
    \IfBeginWith{#1}{http}%
      {\href{#1}{\textcolor{blue}{#2}}}%
      {\href{https://doi.org/#1}{\textcolor{blue}{#2}}}}}
\usepackage{float}
\usepackage{multirow}
\usepackage{array}
\usepackage{booktabs}
\usepackage{colortbl}
\usepackage{tabularx}
\usepackage{longtable}
\usepackage{geometry}
\renewcommand{\arraystretch}{1.6}
\RequirePackage[numbers,sort&compress]{natbib}
\usepackage{hyperref}
\hypersetup{colorlinks=true, linkcolor=red, citecolor=red, urlcolor=magenta}
\numberwithin{equation}{section}
\newcommand{\dd}{\mathrm{d}}
\newcommand{\rH}{r_{h}}
\newcommand{\RH}{R_{h}}
\newcommand{\Ms}{M}
\newcommand{\Rps}{R_{\rm ps}}
\newcommand{\rps}{r_{\rm ps}}
\graphicspath{{figs/}}
\begin{document}
\baselineskip=16pt

\title{\LARGE Simpson-Visser regularization of the quantum Oppenheimer-Snyder-Datt black hole}
\author{Faizuddin Ahmed\,\orcidlink{0000-0003-2196-9622}\\
\small Department of Physics, The Assam Royal Global University, Guwahati, 781035, Assam, India\\
\small \texttt{fahmed@rgu.ac ; faizuddinahmed15@gmail.com}
\and
Ahmad Al-Badawi\,\orcidlink{0000-0002-3127-3453}\\
\small Department of Physics, Al-Hussein Bin Talal University, P. O. Box: 20, 71111, Ma'an, Jordan\\
\small \texttt{ahmadbadawi@ahu.edu.jo}
\and
\.{I}zzet Sakall{\i}\,\orcidlink{0000-0001-7827-9476}\\
\small Physics Department, Eastern Mediterranean University,\\ \small Famagusta 99628, North Cyprus via Mersin 10, Turkiye\\
\small \texttt{izzet.sakalli@emu.edu.tr}}
\date{}
\maketitle

\vspace{-10pt}
\begin{abstract}
\noindent We study the static, spherically symmetric geometry obtained by applying the Simpson-Visser bounce prescription $r\to\sqrt{r^2+a^2}$ to the quantum Oppenheimer-Snyder-Datt (qOSD) black hole of loop quantum cosmology, whose lapse function $A=1-2M/R+\alpha M^2/R^4$ carries the holonomy correction $\alpha$. The two-parameter family interpolates between the qOSD black hole ($a\to0$), the Simpson-Visser black bounce ($\alpha\to0$), and Schwarzschild. The geometry is free of curvature singularities for every $a>0$, the Ricci scalar at the throat is blind to $\alpha$ while the Kretschmann scalar is not, and the null energy condition is violated at the throat by the universal amount $-1/(4\pi a^2)$. The $(\alpha,a)$ plane splits into a two-horizon regular black hole, a single-horizon black hole with a hidden bounce and no Cauchy horizon, and a traversable wormhole; the middle region removes the mass-inflation channel of regular black holes without fine tuning. Because circular orbits are governed by the areal radius alone, the photon sphere, the shadow, the ISCO and the azimuthal epicyclic frequency depend on $\alpha$ only, whereas the Lyapunov exponent and the radial epicyclic frequency carry the exact factor $\sqrt{1-a^2/R^2}$: the shadow constrains only $\alpha$, while the ringdown damping and the twin-peak quasiperiodic-oscillation ratio are sensitive to $a$. We derive the horizon thermodynamics with a first-law entropy that departs from the area law, show that the bounce changes the thermodynamic topological charge from $W=-1$ to $W=0$ at every $\alpha$, and compute scalar, electromagnetic and Dirac quasinormal spectra with third-order WKB, a Chebyshev spectral collocation solver and time-domain integration, which agree with one another and with the eikonal scaling. Greybody bounds, the Hawking emission rate, the sparsity of the cascade and the ringdown quality factor complete the analysis, and in the wormhole region the double-peaked potential produces echoes.
\end{abstract}

Word count: 268.

\vspace{0.2cm}

\small

\section{Introduction}\label{isec1}

Loop quantum gravity replaces the classical collapse of a homogeneous dust ball by a bounce. The quantum Oppenheimer-Snyder-Datt model (qOSD) of Lewandowski, Ma, Yang and Zhang~\cite{Lewandowski:2023} glues the effective loop-quantum-cosmology interior of a collapsing dust cloud~\cite{Ashtekar:2018,Rovelli:2014} to a static exterior, and the junction conditions fix that exterior uniquely: the lapse is $1-2M/R+\alpha M^2/R^4$, where $\alpha=16\sqrt{3}\pi\gamma^3\ell_p^2$ is set by the Barbero-Immirzi parameter and the Planck length. The classical Oppenheimer-Snyder-Datt solution~\cite{Oppenheimer:1939,Datt1938,Datt1999} is recovered when $\alpha\to0$. The $R^{-4}$ correction has attracted a steady stream of follow-up work, on the shadow and linear stability~\cite{Yang:2023,Zhang:2023}, on quasinormal ringing for fields of various spin~\cite{Gong:2024,Skvortsova:2024,Zinhailo:2025}, on lensing, particle tunneling, scalarization and observational bounds~\cite{Luo:2024,Tan:2025,Myung:2025,Chen:2025,Yang:2026,Jiang:2026}, and on the broader class of covariant effective loop black holes~\cite{Konoplya:2025a,Konoplya:2025b,Moreira:2023,Gingrich:2024}; quantum-corrected Schwarzschild geometries of a different origin, derived within the effective-field-theory treatment of gravity, have been analyzed along similar lines~\cite{Battista:2024,Wang:2025}. The exterior is, however, not regular. Written in the areal radius $R$ the Kretschmann scalar diverges at $R=0$, and for $0<\alpha<27M^2/16$ the geometry has an inner horizon at $R_-$ in addition to the event horizon at $R_+$. Both features are inherited from the fact that the qOSD exterior is only meant to hold outside the dust surface; inside, the bouncing dust profile takes over. If one wants to use the exterior as an effective description of the whole spacetime, something has to be done about the origin and about the inner horizon.

Simpson and Visser~\cite{Simpson:2019} proposed a minimal way to do exactly that for Schwarzschild: replace the areal coordinate by $\sqrt{r^2+a^2}$ with $r\in(-\infty,\infty)$. The resulting black bounce is regular for every $a>0$, and depending on whether $a$ is smaller or larger than $2M$ the spacetime is a regular black hole with a bounce hidden behind the horizon, a one-way wormhole with a null throat, or a two-way traversable wormhole. The prescription has since been applied to Reissner-Nordstr\"om, Kerr, AdS, brane-world, $k$-essence and loop-inspired backgrounds~\cite{Franzin:2021,Mazza:2021,Lobo:2021,Lobo:2020,Muniz:2025,Alencar:2025,Alencar:2026,Ahmed:2026a,Ahmed:2026b,AlBadawi:2026}, its field sources have been identified~\cite{Bronnikov:2022}, and its ringing, echoes, lensing, shadows, tidal forces and epicyclic oscillations have all been examined~\cite{Churilova:2020,Yang:2023b,Ou:2022,Tsukamoto:2021a,Nascimento:2020,Guerrero:2021,Stuchlik:2021,Zhou:2020,Bragado:2025,Malik:2025,Lutfuoglu:2026a}. Most of these studies keep a classical seed metric. The suggestion that the same surgery could be applied to the qOSD exterior arose in our own discussion of the problem, and, to our knowledge, the combination of a holonomy-corrected lapse with a Simpson-Visser bounce has not been analyzed in any detail. The present paper fills that gap.

There is a specific reason why the combination is worth more than a routine parameter study. Regular black holes of the Bardeen, Hayward and Ay\'on-Beato-Garc\'ia type~\cite{Bardeen:1968,Hayward:2006,AyonBeato:1998} all carry an inner horizon, and the inner horizon is the seat of the mass-inflation instability first identified by Poisson and Israel~\cite{Poisson:1990}. Carballo-Rubio and collaborators showed that this instability persists for regular cores and that it is not cured by the finiteness of the curvature~\cite{CarballoRubio:2018,CarballoRubio:2021}; ways to evade it require either a degenerate inner horizon~\cite{CarballoRubio:2022}, a nonstandard core~\cite{Bonanno:2021,Eichhorn:2026}, or an asymptotic structure in which the inner horizon is absent altogether. The qOSD black hole has an inner horizon, so it inherits the problem, and the effective loop treatments of Refs.~\cite{Konoplya:2025a,Cao:2024,Zhou:2023} make the same point in related geometries. The bounce provides a third option. If the bounce radius $a$ is placed between $R_-$ and $R_+$, the surface $R=R_-$ is simply not part of the manifold: the areal radius never reaches it, the throat sits at $r=0$ inside the event horizon, and the spacetime has one horizon, no Cauchy horizon, and finite curvature everywhere. We will call this the hidden-bounce regime. It is the only region of the parameter plane in which the qOSD correction and the bounce reinforce one another rather than merely coexisting, and it will be the reference configuration throughout the paper.

The second reason is observational. Static spherically symmetric geometries of the form $-A\,\dd t^2+\dd r^2/A+R(r)^2\dd\Omega^2$ have the property that circular geodesics are governed by $A$ as a function of the areal radius $R$, while the radial part of the metric only enters through derivatives along $r$. As a consequence the photon sphere, the critical impact parameter, the ISCO and the Keplerian frequency are the same as in the seed qOSD metric, whereas the Lyapunov exponent of the photon sphere and the radial epicyclic frequency acquire the exact factor $\sqrt{1-a^2/R^2}$. In observational terms the Event Horizon Telescope shadow~\cite{Akiyama:2019,Akiyama:2022,Vagnozzi:2023} sees $\alpha$ only, whereas the damping time of the ringdown~\cite{Cardoso:2009,Konoplya:2017} and the twin-peak ratio of high-frequency quasiperiodic oscillations~\cite{Stella:1998,Stella:1999,Motta:2014} see $a$. We make this decoupling explicit, quantify it, and use it to organize the observational discussion. It is also the reason why the shadow is not a useful discriminator for this family: over the full physical range of $\alpha$ the critical impact parameter changes by less than four percent, well inside the present Sgr~A$^*$ band.

The third reason is thermodynamic. For a metric with $g_{tt}g_{rr}\neq-1$ in the areal coordinate the surface gravity is not $A'(R_h)/2$; it is $\tfrac12\sqrt{1-a^2/R_h^2}\,A'(R_h)$, so the Hawking temperature is suppressed by the bounce and vanishes when the throat reaches the horizon. Insisting on the first law then forces an entropy that departs from the area law, in the same way as for Bardeen-type solutions~\cite{Ma:2014,Simovic:2024}. We compute this entropy, the heat capacity and the free energy, and we go one step further by applying the topological classification of Wei, Liu and Mann~\cite{Wei:2022,Duan:1979}, which we recently used to map the branch structure of Kalb-Ramond-ModMax black holes~\cite{Sucu:2026}, to the family. The outcome is clean: the topological charge is $W=0$ for every $a>0$, including the Simpson-Visser limit, whereas the Schwarzschild and qOSD seeds have $W=-1$ and $W=0$ respectively. The bounce is therefore a topological transmutation of the Schwarzschild class in the sense of~\cite{Wei:2022,Chen:2026,Xia:2026}.

The paper is organized as follows. Section~\ref{isec2} introduces the metric, verifies its limits, derives the curvature invariants and the effective source, and maps the parameter plane. Section~\ref{isec3} treats the horizon thermodynamics and the thermodynamic topology. Section~\ref{isec4} analyzes null geodesics, the shadow, the Lyapunov exponent and the EHT constraints. Section~\ref{isec5} covers circular timelike orbits, the ISCO and the epicyclic frequencies with their quasiperiodic-oscillation implications, and Sec.~\ref{isec5b} follows a radially infalling observer through the throat and computes the tidal forces. Section~\ref{isec6} derives the master equations for scalar, electromagnetic and Dirac fields and computes the quasinormal spectra by WKB, Chebyshev collocation and time-domain integration, including echoes in the wormhole regime. Section~\ref{isec7} deals with greybody bounds, the Hawking emission rate, sparsity and the ringdown quality factor. Section~\ref{isec8} contains our conclusions. We use $G=c=\hbar=1$ and set $M=1$ in all numerical work; $\alpha$ and $a$ are then measured in units of $M^2$ and $M$.

\section{The bounce-regularized qOSD spacetime}\label{isec2}

We begin with the line element and its two limiting families, and then establish regularity, the source content and the horizon structure. The qOSD exterior in areal coordinates is $\dd s^2=-A(R)\dd t^2+\dd R^2/A(R)+R^2\dd\Omega^2$ with
\begin{equation}
A(R)=1-\frac{2\Ms}{R}+\frac{\alpha\Ms^2}{R^4},\qquad \alpha=16\sqrt{3}\,\pi\gamma^3\ell_p^2 .
\label{eq:qosd}
\end{equation}
In the original derivation $\alpha$ is Planckian and its effect on astrophysical horizons is negligible; in the phenomenological literature~\cite{Yang:2023,Gong:2024,Luo:2024,Yang:2026} it is treated as a free parameter and constrained by observations, and we follow that practice. A word on the scale is in order. With $\gamma\simeq0.2375$ the coefficient is $\alpha\simeq1.17\,\ell_p^2$, so that $\alpha/M^2$ is of order $(\ell_p/M)^2$, which is $10^{-76}$ for a solar-mass black hole and reaches order unity only for Planck-mass remnants. Taking $\alpha/M^2$ in the range $[0,27/16]$, as we do, therefore describes either the final stage of evaporation of a black hole with a fixed holonomy scale, or a phenomenological deformation whose amplitude is left to be bounded by data; both readings are common in the literature and none of the qualitative conclusions below depends on which is adopted, since the bounce parameter $a$ is in any case a free length that the qOSD model does not fix. The Simpson-Visser regularization consists of the replacement $R\to\sqrt{r^2+a^2}$ with $r\in(-\infty,\infty)$ in the areal function only, together with $\dd R^2/A\to\dd r^2/A$. The metric we study is therefore
\begin{equation}
\dd s^{2}=-A(r)\,\dd t^{2}+\frac{\dd r^{2}}{A(r)}+(r^2+a^2)\left(\dd\theta^2+\sin^2\theta\,\dd\phi^2\right),
\label{eq:metric}
\end{equation}
with
\begin{equation}
A(r)=1-\frac{2\Ms}{\sqrt{r^2+a^2}}+\frac{\alpha\Ms^2}{(r^2+a^2)^2}.
\label{eq:lapse}
\end{equation}
Throughout we follow the black-hole literature and call $A$ the lapse function, although in the ADM sense the lapse is $N=\sqrt{A}$; nothing below depends on this distinction, since every quantity we compute is built from $g_{tt}=-A$ and $g_{rr}=1/A$.
Written back in the areal radius $R=\sqrt{r^2+a^2}\geq a$ the same geometry reads
\begin{equation}
\dd s^{2}=-A(R)\,\dd t^{2}+\frac{\dd R^{2}}{A(R)\left(1-a^2/R^2\right)}+R^2\dd\Omega^2,
\label{eq:arealform}
\end{equation}
which makes the role of $a$ transparent: it does not touch $g_{tt}$, it multiplies $g_{RR}$ by the factor $(1-a^2/R^2)^{-1}$, and it truncates the range of the areal radius at the throat $R=a$. Two copies of the region $R\geq a$ are glued at the throat, one for $r>0$ and one for $r<0$. The three limits are immediate and were verified symbolically: $a\to0$ returns the qOSD lapse~(\ref{eq:qosd}); $\alpha\to0$ returns the Simpson-Visser black bounce $A=1-2M/\sqrt{r^2+a^2}$~\cite{Simpson:2019}; and $a\to0$ together with $\alpha\to0$ returns Schwarzschild. Every result below reduces to the known ones in these limits, and we shall point this out at each step.

\begin{figure}[!htbp]
\centering
\includegraphics[width=\textwidth]{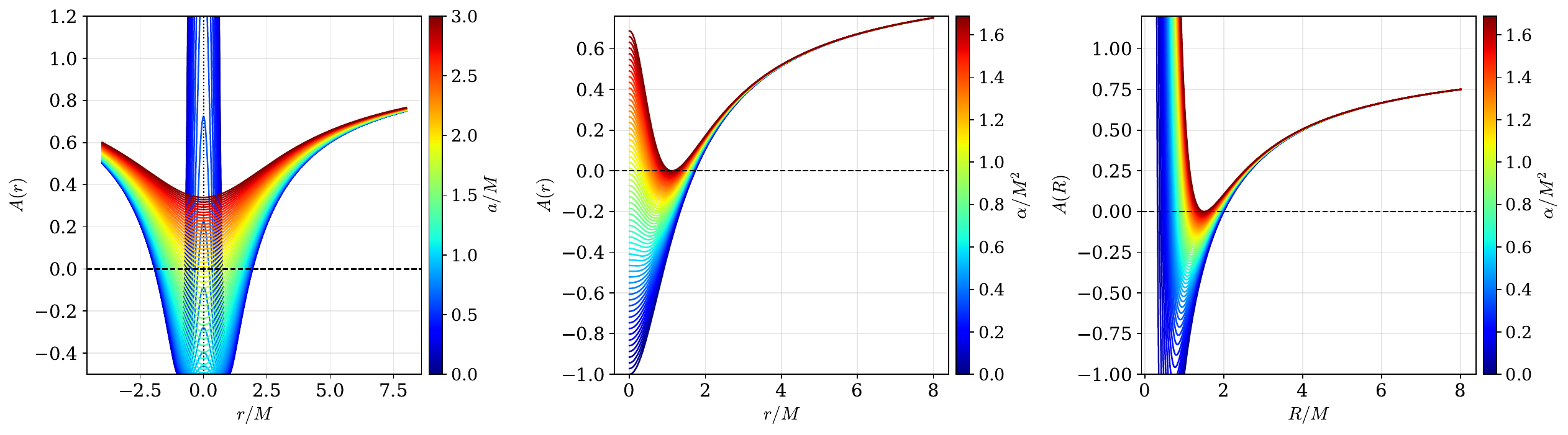}
\caption{Left: the lapse $A(r)$ of Eq.~(\ref{eq:lapse}) at $\alpha=0.5M^2$ as the bounce parameter $a$ runs from $0$ to $3M$ (color bar), shown for $A>-0.5$. The function is even in $r$; a zero of $A$ at $r_h>0$ is an event horizon, and curves that stay positive describe traversable wormholes. Middle: $A(r)$ at $a=M$ as $\alpha$ runs from $0$ to the extremal value $27M^2/16$, with the vertical range cut at $A=0.76$ to display the horizon region. Right: the same family written in the areal radius $R$, where the bounce is invisible and the lapse coincides with the qOSD one. The dashed line marks $A=0$.}
\label{fig:lapse}
\end{figure}

Figure~\ref{fig:lapse} shows the lapse in the two coordinates. In the areal coordinate the curves are those of the qOSD black hole: for $\alpha<\alpha_{\rm ext}$ the function crosses zero twice, at $R_-$ and $R_+$, and for $\alpha>\alpha_{\rm ext}$ it stays positive. The extremal value follows from the double-root condition on $R^4-2MR^3+\alpha M^2=0$, that is $R^4-2MR^3+\alpha M^2=0$ together with $4R^3-6MR^2=0$, which gives
\begin{equation}
R_{\rm ext}=\frac{3M}{2},\qquad \alpha_{\rm ext}=\frac{27M^2}{16}\simeq1.6875\,M^2 .
\label{eq:ext}
\end{equation}
For small $\alpha$ the two roots behave as $R_+\simeq2M-\alpha/(8M)$ and $R_-\simeq(\alpha M/2)^{1/3}$, and at $\alpha=M^2$ the quartic factorizes as $(R-M)(R^3-MR^2-M^2R-M^3)$, so that the inner horizon sits exactly at $R_-=M$; this coincidence is visible in Table~\ref{tab:regimes}. In the bounce coordinate the horizons are $r_\pm=\sqrt{R_\pm^2-a^2}$ and they exist only when $R_\pm>a$. As $a$ increases at fixed $\alpha$, the left panel of Fig.~\ref{fig:lapse} shows the inner horizon disappearing first, when $a$ crosses $R_-$, and the event horizon disappearing second, when $a$ crosses $R_+$. Beyond that point $A$ is positive on the whole real line, with a minimum $A(0)=A(R=a)$ at the throat.

\begin{figure}[!htbp]
\centering
\includegraphics[width=0.62\textwidth]{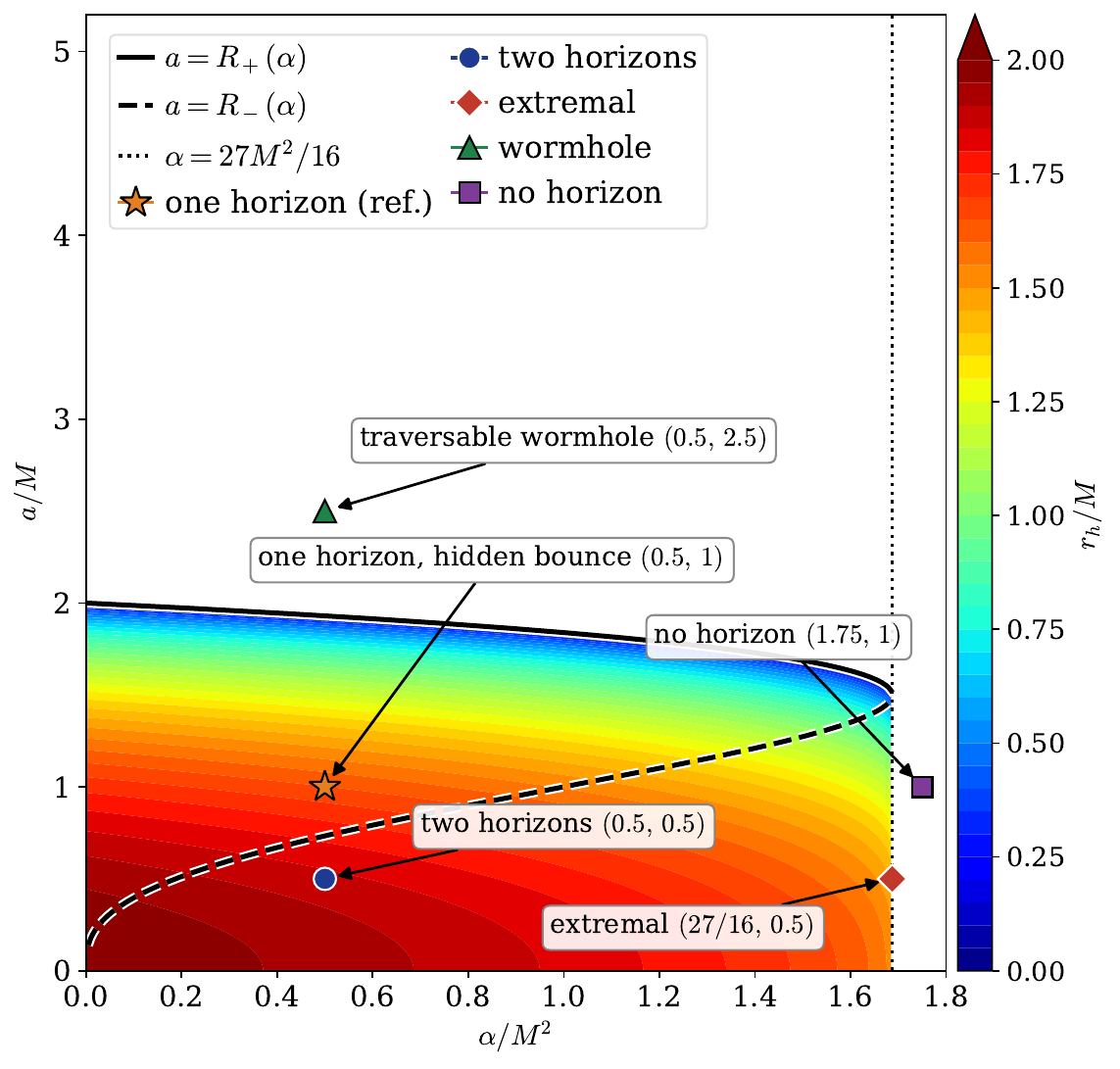}
\caption{Parameter plane of the bounce-qOSD family. The color scale gives the event-horizon position $r_h=\sqrt{R_+^2-a^2}$ in the bounce coordinate; the white region has no horizon. The solid and dashed black curves are $a=R_+(\alpha)$ and $a=R_-(\alpha)$ and separate the traversable-wormhole region above $R_+$, the hidden-bounce region with a single horizon between the two curves, and the two-horizon region below $R_-$. The dotted vertical line is $\alpha_{\rm ext}$. Each marker is one representative configuration, tagged by an arrow with its causal type and its $(\alpha/M^2,a/M)$ coordinates: the reference point $(0.5,1)$ with a single horizon (star), a two-horizon black hole at $(0.5,0.5)$ (circle), an extremal configuration at $(27/16,0.5)$ (diamond), a traversable wormhole at $(0.5,2.5)$ (triangle), and a horizonless over-extremal geometry at $(1.75,1)$ (square).}
\label{fig:phase}
\end{figure}

The phase diagram of Fig.~\ref{fig:phase} summarizes this. Three regions appear. For $a<R_-(\alpha)$ the spacetime has two horizons in $r$, the throat $r=0$ lies inside the inner horizon where $A>0$ and $r$ is spacelike, and the causal structure is that of a Reissner-Nordstr\"om-like regular black hole with a wormhole throat replacing the timelike singularity; we call this the two-horizon region. For $R_-(\alpha)<a<R_+(\alpha)$ there is a single horizon, the throat lies between the horizon and what would have been the inner horizon, and $A(r=0)<0$; in this hidden-bounce region the coordinate $r$ is timelike at the throat, the bounce is a spacelike surface crossed by every infalling observer, and there is no Cauchy horizon. For $a>R_+(\alpha)$ the spacetime is a horizonless traversable wormhole with $A>0$ everywhere. The curve $a=R_+(\alpha)$ is a one-way wormhole with a null throat, and $a=R_-(\alpha)$ is the locus where the inner horizon degenerates with the throat. The region above $\alpha_{\rm ext}$ is a wormhole for every $a>0$ and a naked singularity at $a=0$; on the line $\alpha=\alpha_{\rm ext}$ with $a<3M/2$ the two horizons coincide at $R=3M/2$, the temperature vanishes, and the bounce sits at $r_h=\sqrt{9M^2/4-a^2}$ behind a degenerate horizon. Figure~\ref{fig:phase} marks one representative of each of these cases, and Table~\ref{tab:regimes} includes the extremal and over-extremal rows. A useful way to read the diagram is to follow a vertical line at fixed $\alpha$ upward, which is what an evaporating black hole would do if $a$ stayed fixed while $M$ decreased, since $a/M$ and $\alpha/M^2$ both grow along the evaporation. Such a black hole starts deep in the two-horizon region, loses its inner horizon when $a/M$ crosses $R_-/M$, and finally loses its event horizon when $a/M$ reaches $R_+/M$, at which point, as we show in Sec.~\ref{isec3}, its temperature has dropped to zero and evaporation stops; the endpoint is a one-way wormhole rather than a naked singularity or a Planck-mass remnant with divergent curvature. The alternative endpoint, extremality at $\alpha_{\rm ext}$ with $a$ still below $R_+$, is reached instead when the vertical line hits the dotted boundary first, which happens for $a/M$ below about $1.5$ at the moment $\alpha/M^2$ passes $27/16$. The reference point $(\alpha,a)=(0.5M^2,M)$ sits in the hidden-bounce region, where $R_+=1.9305M$, $R_-=0.7336M$, $r_h=1.6513M$, and we use it for all single-parameter-set figures. Table~\ref{tab:regimes} collects the horizon data on a grid.

{\setlength{\tabcolsep}{12pt}\renewcommand{\arraystretch}{1.6}\small
\begin{longtable}{c c c c c c c c}
\caption{Horizon structure of the bounce-qOSD spacetime for a grid of $(\alpha,a)$ with $M=1$, including the extremal value $\alpha_{\rm ext}=27/16$, where $R_+=R_-=3/2$ and $T_H=0$, and an over-extremal value $\alpha=1.75$, which is a naked singularity at $a=0$ and a horizonless wormhole for $a>0$. $R_\pm$ are the areal horizon radii (roots of $A(R)=0$), $r_h$ and $r_{\rm in}$ the corresponding values of the bounce coordinate, and $T_H$ the Hawking temperature of the outer horizon.}\label{tab:regimes}\\
\cellcolor{brown!30}$\alpha/M^2$ & \cellcolor{brown!30}$a/M$ & \cellcolor{brown!30}$R_+/M$ & \cellcolor{brown!30}$R_-/M$ & \cellcolor{brown!30}$r_h/M$ & \cellcolor{brown!30}$r_{\rm in}/M$ & \cellcolor{brown!30}regime & \cellcolor{brown!30}$T_H M$ \\ \hline\hline
\endfirsthead
\cellcolor{brown!30}$\alpha/M^2$ & \cellcolor{brown!30}$a/M$ & \cellcolor{brown!30}$R_+/M$ & \cellcolor{brown!30}$R_-/M$ & \cellcolor{brown!30}$r_h/M$ & \cellcolor{brown!30}$r_{\rm in}/M$ & \cellcolor{brown!30}regime & \cellcolor{brown!30}$T_H M$ \\ \hline\hline
\endhead
\hline\hline
\endfoot
0.25 & 0.00 & 1.9672 & 0.5575 & 1.9672 & 0.5575 & two horizons & 0.03843 \\
0.25 & 0.50 & 1.9672 & 0.5575 & 1.9026 & 0.2467 & two horizons & 0.03716 \\
0.25 & 1.00 & 1.9672 & 0.5575 & 1.6940 & -- & one horizon & 0.03309 \\
0.25 & 1.50 & 1.9672 & 0.5575 & 1.2727 & -- & one horizon & 0.02486 \\
0.25 & 2.00 & 1.9672 & 0.5575 & -- & -- & wormhole & -- \\
0.25 & 2.50 & 1.9672 & 0.5575 & -- & -- & wormhole & -- \\
0.50 & 0.00 & 1.9305 & 0.7336 & 1.9305 & 0.7336 & two horizons & 0.03677 \\
0.50 & 0.50 & 1.9305 & 0.7336 & 1.8646 & 0.5368 & two horizons & 0.03551 \\
0.50 & 1.00 & 1.9305 & 0.7336 & 1.6513 & -- & one horizon & 0.03145 \\
0.50 & 1.50 & 1.9305 & 0.7336 & 1.2153 & -- & one horizon & 0.02315 \\
0.50 & 2.00 & 1.9305 & 0.7336 & -- & -- & wormhole & -- \\
0.50 & 2.50 & 1.9305 & 0.7336 & -- & -- & wormhole & -- \\
1.00 & 0.00 & 1.8393 & 1.0000 & 1.8393 & 1.0000 & two horizons & 0.03192 \\
1.00 & 0.50 & 1.8393 & 1.0000 & 1.7700 & 0.8660 & two horizons & 0.03072 \\
1.00 & 1.00 & 1.8393 & 1.0000 & 1.5437 & 0.0000 & two horizons & 0.02679 \\
1.00 & 1.50 & 1.8393 & 1.0000 & 1.0644 & -- & one horizon & 0.01847 \\
1.00 & 2.00 & 1.8393 & 1.0000 & -- & -- & wormhole & -- \\
1.00 & 2.50 & 1.8393 & 1.0000 & -- & -- & wormhole & -- \\
1.50 & 0.00 & 1.6883 & 1.2732 & 1.6883 & 1.2732 & two horizons & 0.02103 \\
1.50 & 0.50 & 1.6883 & 1.2732 & 1.6126 & 1.1709 & two horizons & 0.02008 \\
1.50 & 1.00 & 1.6883 & 1.2732 & 1.3603 & 0.7880 & two horizons & 0.01694 \\
1.50 & 1.50 & 1.6883 & 1.2732 & 0.7748 & -- & one horizon & 0.00965 \\
1.50 & 2.00 & 1.6883 & 1.2732 & -- & -- & wormhole & -- \\
1.50 & 2.50 & 1.6883 & 1.2732 & -- & -- & wormhole & -- \\
1.6875 & 0.00 & 1.5000 & 1.5000 & 1.5000 & 1.5000 & extremal & 0 \\
1.6875 & 0.50 & 1.5000 & 1.5000 & 1.4142 & 1.4142 & extremal & 0 \\
1.6875 & 1.00 & 1.5000 & 1.5000 & 1.1180 & 1.1180 & extremal & 0 \\
1.6875 & 2.00 & 1.5000 & 1.5000 & -- & -- & wormhole & -- \\
1.7500 & 0.00 & -- & -- & -- & -- & naked singularity & -- \\
1.7500 & 0.50 & -- & -- & -- & -- & wormhole & -- \\
1.7500 & 1.00 & -- & -- & -- & -- & wormhole & -- \\
1.7500 & 2.00 & -- & -- & -- & -- & wormhole & -- \\
\end{longtable}}

The regularity of the metric is established from the curvature invariants. For a metric of the form~(\ref{eq:metric}) with areal function $R(r)$ the independent components of the Riemann tensor in an orthonormal frame are the four sectional curvatures $E_1=-A''/2$, $E_2=-A'R'/(2R)$, $E_3=-(AR''+A'R'/2)/R$ and $E_4=(1-AR'^2)/R^2$, where primes denote $r$ derivatives, and the Ricci scalar and Kretschmann scalar are $\mathcal{R}=2(E_1+2E_2+2E_3+E_4)$ and $\mathcal{K}=4(E_1^2+2E_2^2+2E_3^2+E_4^2)$. We verified that these expressions return $\mathcal{R}=0$ and $\mathcal{K}=48M^2/r^6$ for Schwarzschild and $\mathcal{K}=4(9M^2-8aM+3a^2)/a^6$ at the throat of the Simpson-Visser spacetime, in agreement with~\cite{Simpson:2019}. For the full family the two invariants at the throat are
\begin{align}
\mathcal{R}\big|_{r=0}&=\frac{2(3M-a)}{a^3},
\label{eq:ricci0}\\
\mathcal{K}\big|_{r=0}&=\frac{4}{a^{12}}\Big[M^2\big(2M\alpha-a^3\big)^2+a^8+2\big(M^2\alpha-2Ma^3+a^4\big)^2\Big].
\label{eq:kret0}
\end{align}
Both are finite for every $a>0$, and since the invariants are smooth functions of $r$ elsewhere, the geometry has no curvature singularity. Equation~(\ref{eq:ricci0}) has a feature worth noting: the Ricci scalar at the throat does not depend on $\alpha$. The reason is that $A'(r)$ vanishes at $r=0$ for any $A(R)$, so $E_2=0$ there, and the surviving combination $-A''(0)-4A(a)/a^2+2/a^2$ cancels the $\alpha$ terms identically once $A''(0)=A'(R{=}a)/a$ is used. The Kretschmann scalar retains the $\alpha$ dependence through the squares. In the Simpson-Visser limit Eq.~(\ref{eq:kret0}) reduces to the value quoted above, and it grows as $a^{-6}$ for $a\to0$, which is how the qOSD singularity is recovered in the limit; the extra $\alpha$-dependent terms scale as $a^{-12}$, one power of $a^{-6}$ faster, reflecting the stronger $R^{-4}$ falloff of the correction.

\begin{figure}[!htbp]
\centering
\includegraphics[width=\textwidth]{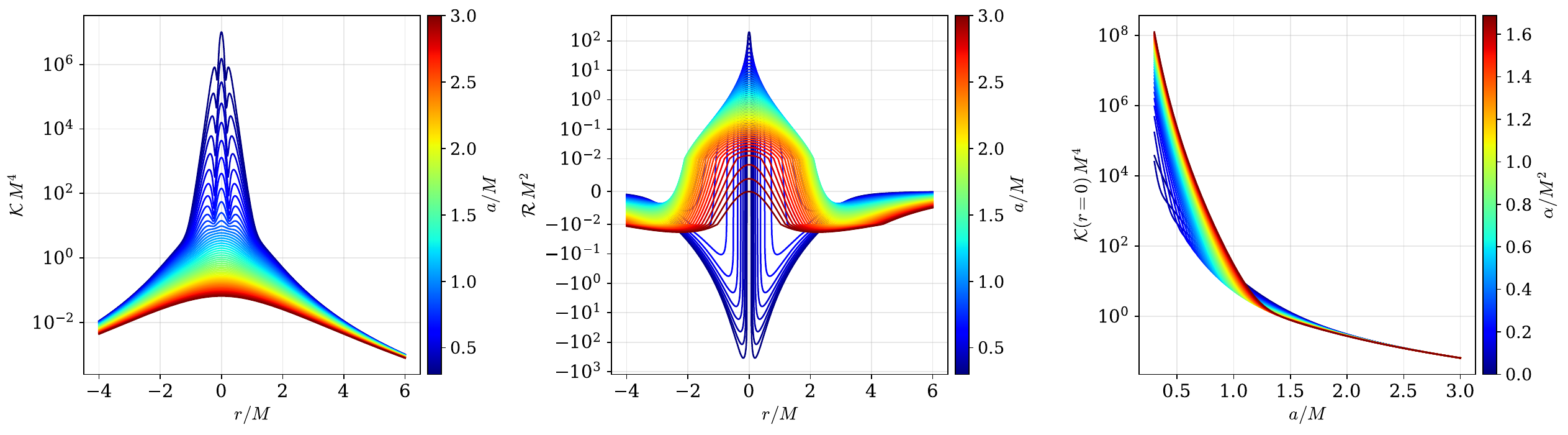}
\caption{Left: Kretschmann scalar along the bounce coordinate at $\alpha=0.5M^2$ for $a$ from $0.3M$ to $3M$. Middle: Ricci scalar on the same grid, on a symmetric logarithmic scale. Right: Kretschmann scalar at the throat, Eq.~(\ref{eq:kret0}), as a function of $a$ for $\alpha$ from $0$ to $\alpha_{\rm ext}$. All curves are finite at $r=0$, and the throat value is dominated by the $a^{-12}$ term for $a\lesssim M$.}
\label{fig:curv}
\end{figure}

Figure~\ref{fig:curv} shows the two invariants as ribbons in $a$. The Kretschmann profile is symmetric in $r$, peaks at the throat, and its throat value decreases by six orders of magnitude between $a=0.3M$ and $a=3M$. The Ricci scalar changes sign at $a=3M$, as Eq.~(\ref{eq:ricci0}) says, and is positive at the throat of all the black-hole configurations since those have $a<R_+<2M$.

We next ask what matter, in Einstein's equations $G_{\mu\nu}=8\pi T_{\mu\nu}$, would support the geometry; the action and field equations that reproduce the answer exactly, in terms of a phantom scalar and a magnetic nonlinear electrodynamics, are given in Appendix~\ref{isecA}. With $T^\mu{}_\nu={\rm diag}(-\rho,p_r,p_t,p_t)$ the sectional curvatures give $8\pi\rho=2E_3+E_4$, $8\pi p_r=-(2E_2+E_4)$ and $8\pi p_t=-(E_1+E_2+E_3)$. At the throat we find
\begin{equation}
\rho\big|_{r=0}=\frac{4Ma^3-2\alpha M^2-a^4}{8\pi a^6},\qquad
p_t\big|_{r=0}=\frac{a^4-Ma^3-\alpha M^2}{8\pi a^6},\qquad
\rho+p_r\big|_{r=0}=-\frac{A(R{=}a)}{4\pi a^2},
\label{eq:source0}
\end{equation}
where the last relation was obtained by direct simplification and holds for any lapse. It shows that the sign of $\rho+p_r$ at the throat tracks the sign of $A(a)$, and hence differs between the three regions of Fig.~\ref{fig:phase}. The covariant statement is cleaner. For a radial null vector $k^\mu=(1/A,\pm1,0,0)$ one has $T_{\mu\nu}k^\mu k^\nu=(\rho+p_r)/A$, and at the throat this equals
\begin{equation}
T_{\mu\nu}k^\mu k^\nu\big|_{r=0}=-\frac{1}{4\pi a^2},
\label{eq:nec0}
\end{equation}
independently of $\alpha$, of $M$, and of which side of the horizon the throat is on. The null energy condition is therefore violated at the throat of every member of the family by a fixed, purely geometric amount, in the same way as for the original black bounce~\cite{Simpson:2019,Lobo:2021} and for its field-theoretic sources~\cite{Bronnikov:2022}; the general analysis of energy conditions in static spherically symmetric effective geometries of~\cite{Wang:2026} covers this case. The holonomy correction cannot repair this; it only changes how fast the violation decays away from the throat. For the tangential null vector one finds $T_{\mu\nu}k^\mu k^\nu=\rho+p_t$, without the $1/A$ factor.

\begin{figure}[!htbp]
\centering
\includegraphics[width=0.92\textwidth]{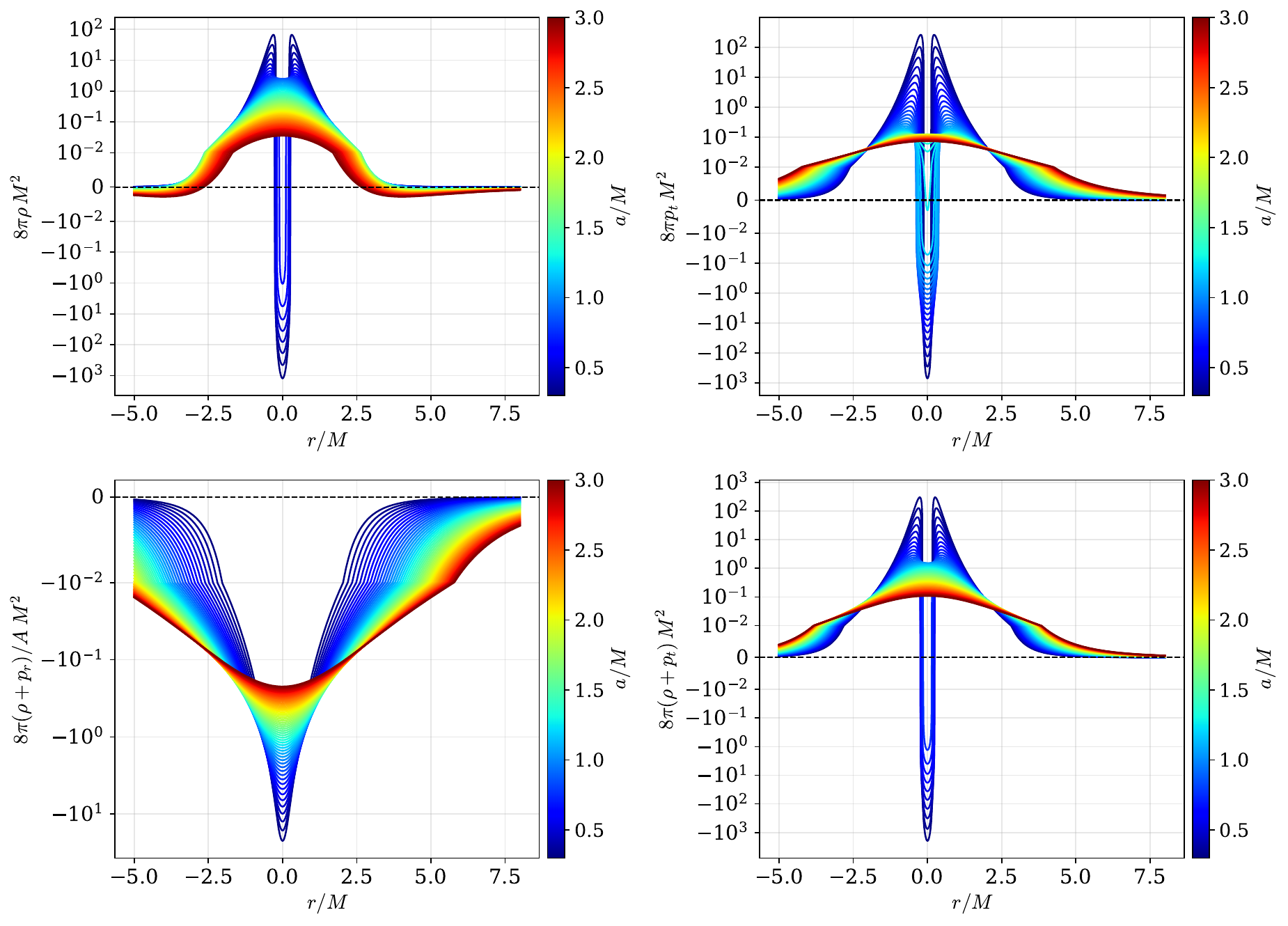}
\caption{Effective source of the geometry at $\alpha=0.5M^2$ for $a$ from $0.3M$ to $3M$: energy density (top left), tangential pressure (top right), the radial null-energy quantity $(\rho+p_r)/A$ (bottom left), and $\rho+p_t$ (bottom right), all on symmetric logarithmic scales. The radial null-energy quantity is negative at the throat with the universal value $-1/(4\pi a^2)$ of Eq.~(\ref{eq:nec0}) and approaches zero from below at large $|r|$.}
\label{fig:src}
\end{figure}

Figure~\ref{fig:src} displays $\rho$, $p_t$ and the two null-energy quantities. Far from the throat all four decay as inverse powers of $|r|$. The leading tail of the density is $\rho\simeq-a^2/(8\pi r^4)+Ma^2/(2\pi r^5)$, which comes entirely from the bounce and is negative, exactly as in the Simpson-Visser case, while the holonomy correction first appears at order $r^{-6}$ with the positive coefficient $3\alpha M^2/(8\pi)$; the radial combination behaves as $\rho+p_r\simeq-a^2/(4\pi r^4)$. The radial null-energy quantity is negative on the whole axis, reaches its most negative value at the throat, and its magnitude there decreases monotonically with $a$. The tangential combination $\rho+p_t$ is positive close to the throat for the smaller values of $a$ and negative for the larger ones, so the tangential null energy condition is not violated universally; the details can be read off the bottom-right panel.

\begin{figure}[!htbp]
\centering
\includegraphics[width=0.95\textwidth]{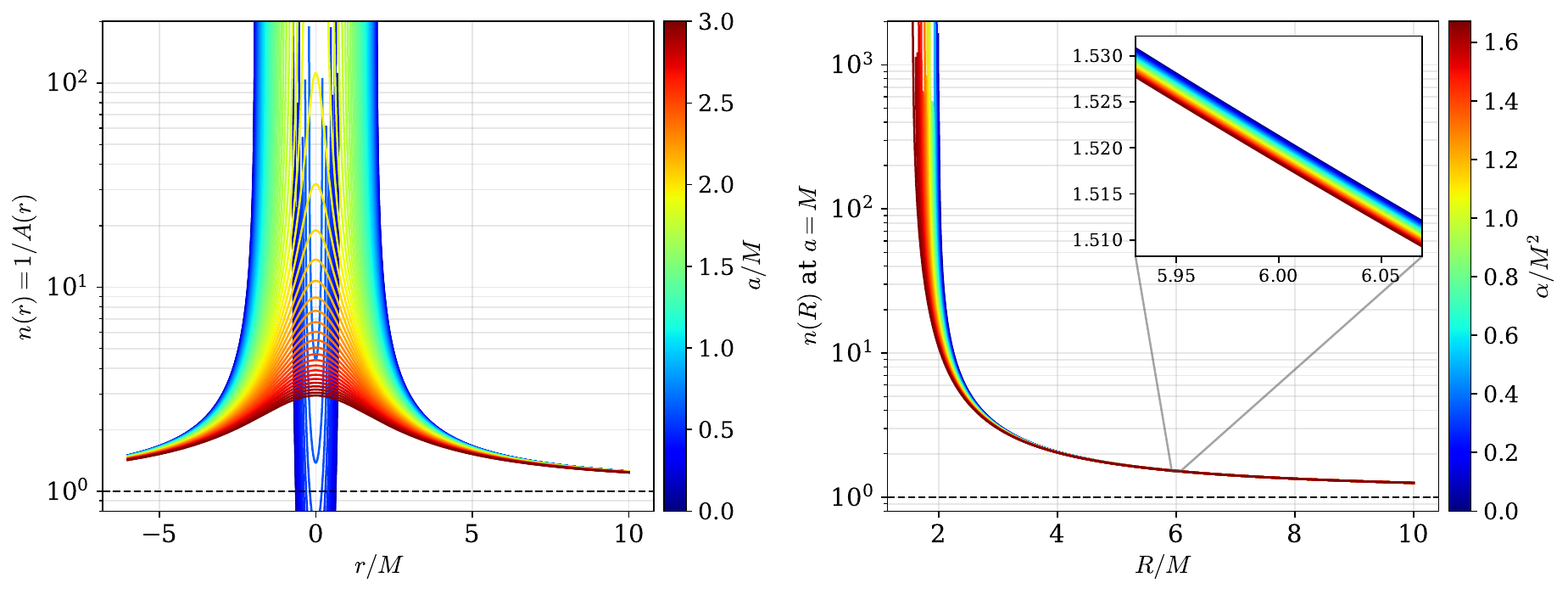}
\caption{Effective refractive index of the geometry. Left: $n(r)$ of Eq.~(\ref{eq:nr}) at $\alpha=0.5M^2$ for $a$ from $0$ to $3M$, plotted where $A>0$; black-hole configurations diverge at $r=\pm r_h$, wormholes are finite through the throat with a maximum $1/A(a)$ at $r=0$. Right: $n(R)$ of Eq.~(\ref{eq:nR}) at $a=M$ for $\alpha$ from $0$ to $\alpha_{\rm ext}$, outside the horizon; the divergence at $R=R_+(\alpha)$ moves inward as $\alpha$ grows, and the inset at the upper right magnifies $5.93\leq R/M\leq6.07$ by about twenty times, with guide lines, to show the ordering of the curves in $\alpha$ at intermediate radius. Both panels are on a logarithmic scale and the dashed line is $n=1$.}
\label{fig:refr}
\end{figure}

The radial propagation of light gives a compact way to see how the two parameters enter the causal structure, and it makes contact with the optical description of the Simpson-Visser model. Setting $\dd s^2=0$ with $\dd\theta=\dd\phi=0$ in Eq.~(\ref{eq:arealform}) gives $\dd R/\dd t=\pm A(R)\sqrt{1-a^2/R^2}$, so the coordinate speed of light in the areal radius and the corresponding effective refractive index are
\begin{equation}
c(R)=A(R)\sqrt{1-\frac{a^2}{R^2}},\qquad
n(R)=\frac{1}{c(R)}=\left(1-\frac{2\Ms}{R}+\frac{\alpha\Ms^2}{R^4}\right)^{-1}\left(1-\frac{a^2}{R^2}\right)^{-1/2},
\label{eq:nR}
\end{equation}
whereas in the bounce coordinate the same rays obey $\dd r/\dd t=\pm A(r)$ and
\begin{equation}
n(r)=\frac{1}{|\dd r/\dd t|}=\left(1-\frac{2\Ms}{\sqrt{r^2+a^2}}+\frac{\alpha\Ms^2}{(r^2+a^2)^2}\right)^{-1}.
\label{eq:nr}
\end{equation}
At $\alpha=0$ Eq.~(\ref{eq:nR}) is the refractive index of the Simpson-Visser spacetime \cite{Simpson:2019}, and the holonomy correction adds the $R^{-4}$ term inside the first factor. The two expressions describe the same physics and differ only by the Jacobian $\dd R/\dd r=\sqrt{1-a^2/R^2}$. The factor $(1-a^2/R^2)^{-1/2}$ in Eq.~(\ref{eq:nR}) diverges at the throat for every member of the family, but this is a coordinate effect: $R$ has a minimum at $r=0$, so $\dd R/\dd t$ must vanish there whatever the geometry, and Eq.~(\ref{eq:nr}) shows that the physical coordinate speed $|\dd r/\dd t|=|A(a)|$ at the throat is finite and nonzero except on the boundary $a=R_+(\alpha)$. The index $n(r)$ is even in $r$, tends to $1+2\Ms/|r|+O(r^{-2})$ at large distance, diverges at the horizons $r=\pm\rH$ of the black-hole configurations, where light freezes in coordinate time, and for the wormholes is finite everywhere with its maximum $n(0)=1/A(a)$ at the throat, which grows without bound as $a\to R_+(\alpha)^+$. Since $R_+\simeq2\Ms-\alpha/(8\Ms)$, the black-hole-to-wormhole transition, and with it the divergence of the throat index, occurs at a slightly smaller bounce than in the Simpson-Visser case. Figure~\ref{fig:refr} shows both forms of the index. The optical geometry also explains the shadow result of Sec.~\ref{isec4} in advance: light rays are geodesics of the optical metric $\dd l^2=n^2\dd r^2+n\,(r^2+a^2)\,\dd\Omega^2$, whose circular orbits sit where $(A/R^2)'=0$ in the areal radius, so the photon sphere and the shadow depend on $\alpha$ through Eqs.~(\ref{eq:ps}) and~(\ref{eq:bc}) and not on $a$, in contrast with the Simpson-Visser value $b_c=3\sqrt3\Ms$, which is fixed for all $a$.

The asymptotic structure is that of the seed metric. Expanding Eq.~(\ref{eq:lapse}) at large $r$ gives $A=1-2M/r+Ma^2/r^3+\alpha M^2/r^4+O(r^{-5})$, so the ADM mass is $M$, the bounce corrections start at $r^{-3}$, and the holonomy correction at $r^{-4}$. Neither modifies the post-Newtonian sector at first order, which is why weak-field tests are insensitive to both parameters and one must go to the strong-field observables of the following sections.
\section{Horizon thermodynamics and thermodynamic topology}\label{isec3}

We now turn the horizon data into thermodynamic quantities, taking care of the factor by which the bounce modifies the surface gravity, and then classify the resulting family with the topological method of Wei, Liu and Mann. The areal form~(\ref{eq:arealform}) has $g_{tt}g_{RR}\neq-1$, so the surface gravity is not $A'(\RH)/2$ but $\kappa=\tfrac12\sqrt{-g^{tt}g^{RR}}\,|\partial_R g_{tt}|$ evaluated at the horizon, which gives $\kappa=\tfrac12\sqrt{1-a^2/\RH^2}\,A'(\RH)$. The same result follows in the bounce coordinate from $\kappa=A'(\rH)/2$ with $\dd A/\dd r=(r/R)\,\dd A/\dd R$. Solving $A(\RH)=0$ for the mass gives
\begin{equation}
\Ms(\RH)=\frac{\RH^3}{\alpha}\left[1-\sqrt{1-\frac{\alpha}{\RH^2}}\right]\;\xrightarrow{\;\alpha\to0\;}\;\frac{\RH}{2}+\frac{\alpha}{8\RH}+\dots,
\label{eq:MofR}
\end{equation}
which requires $\RH^2\geq\alpha$; the branch with the minus sign is the one connected to Schwarzschild, and we checked the limit symbolically. The Hawking temperature is then
\begin{equation}
T_H=\frac{\kappa}{2\pi}=\sqrt{1-\frac{a^2}{\RH^2}}\;\frac{\Ms}{2\pi\RH^2}\left(1-\frac{2\alpha\Ms}{\RH^3}\right),
\label{eq:TH}
\end{equation}
with $\Ms=\Ms(\RH)$. It vanishes at two places: at $\RH=a$, where the throat reaches the horizon and the black hole turns into a one-way wormhole, and at $\RH=R_0\equiv\sqrt{4\alpha/3}$, where $2\alpha\Ms/\RH^3=1$ and the two areal horizons merge. The second condition is the fixed-$\alpha$ version of Eq.~(\ref{eq:ext}), as one checks by inserting $\Ms_{\rm ext}=4\sqrt{\alpha/27}$. The physical branch is $\RH>\max(a,R_0)$, and below $R_0$ the function~(\ref{eq:TH}) becomes negative, which is the signature of the inner-horizon branch. For $a=\alpha=0$ Eq.~(\ref{eq:TH}) reduces to $1/(8\pi\Ms)$, and for $\alpha=0$ it reproduces the Simpson-Visser temperature $\sqrt{1-a^2/4\Ms^2}/(8\pi\Ms)$~\cite{Ditta:2024,AlBadawi:2026}.

The entropy is defined by the first law $\dd\Ms=T_H\,\dd S$ at fixed $\alpha$ and $a$, which is the standard choice for regular black holes~\cite{Ma:2014,Simovic:2024} once the parameters are treated as constants rather than as thermodynamic variables. Differentiating $A(\RH,\Ms(\RH))=0$ gives $\dd\Ms/\dd\RH=A'(\RH)/[\tfrac{2}{\RH}(1-\alpha\Ms/\RH^3)]$, and dividing by Eq.~(\ref{eq:TH}) yields
\begin{equation}
S(\RH)=\int_{R_{\rm min}}^{\RH}\frac{2\pi R\,\dd R}{\left(1-\alpha\Ms(R)/R^3\right)\sqrt{1-a^2/R^2}},\qquad R_{\rm min}=\max(a,R_0),
\label{eq:S}
\end{equation}
where the lower limit is the zero-temperature configuration and we set $S=0$ there. For $\alpha=0$ the integral is elementary,
\begin{equation}
S_{\alpha=0}=\pi\left[\RH\sqrt{\RH^2-a^2}+a^2\ln\frac{\RH+\sqrt{\RH^2-a^2}}{a}\right],
\label{eq:SSV}
\end{equation}
which reduces to the area law $\pi\RH^2$ as $a\to0$ and exceeds it for $a>0$ because of the logarithm. For $\alpha>0$ the factor $(1-\alpha\Ms/R^3)^{-1}$ increases the entropy further, and we evaluate the integral numerically. The heat capacity $C=\dd\Ms/\dd T_H=(\dd\Ms/\dd\RH)/(\dd T_H/\dd\RH)$ and the free energy $F=\Ms-T_HS$ follow.

\begin{figure}[!htbp]
\centering
\includegraphics[width=0.92\textwidth]{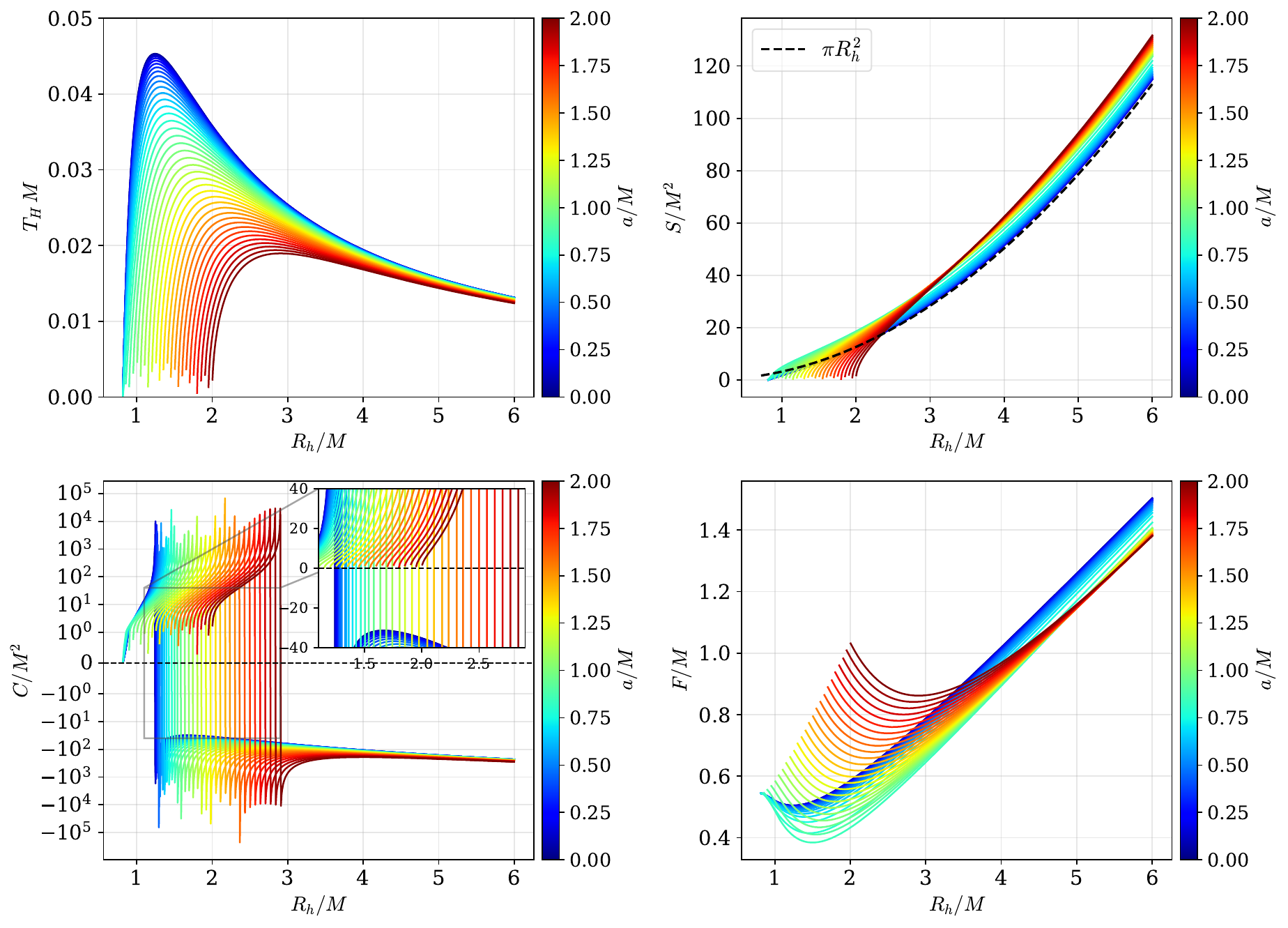}
\caption{Thermodynamics of the bounce-qOSD black hole at $\alpha=0.5M^2$ as functions of the areal horizon radius, for $a$ from $0$ to $2M$: Hawking temperature (top left), first-law entropy compared with the area law $\pi R_h^2$ (dashed, top right), heat capacity on a symmetric logarithmic scale (bottom left), with a linear-scale inset that magnifies the window $1.1\leq R_h/M\leq2.9$, $|C|\leq40M^2$ where the curves diverge and change sign at their Davies points, and free energy (bottom right). Each ribbon starts at the zero-temperature radius $R_{\rm min}=\max(a,R_0)$.}
\label{fig:thermo}
\end{figure}

Figure~\ref{fig:thermo} and Table~\ref{tab:thermo} contain the results. The temperature rises from zero at $R_{\rm min}$, reaches a maximum, and then decays as $1/\RH$; the maximum moves outward and downward as $a$ grows, from $\RH=1.24M$ at $a=0$ to $\RH=2.77M$ at $a=1.9M$. At the reference point the temperature is $T_H=0.0315/M$, about $21\%$ below the Schwarzschild value for the same mass, of which $8\%$ is due to $\alpha$ and the rest to the bounce factor $\sqrt{1-a^2/\RH^2}=0.855$. The entropy exceeds the area law throughout, by $8\%$ at $\RH=2M$ for $a=M/2$ and by $42\%$ for $a=M$, and the excess grows as $\RH\to R_{\rm min}$, where the integrand has an integrable inverse-square-root singularity. The heat capacity is positive on the small-black-hole branch between $R_{\rm min}$ and the temperature maximum, diverges there, and is negative on the large branch, so the family has a Davies point at every $a$. The free energy is monotonic in $\RH$ on the large branch, and in the absence of a thermal bath there is no Hawking-Page-type transition; the wormhole configurations with $\RH<a$ are not part of the thermodynamic family at all, since they have no horizon.

It is worth comparing the two sources of the entropy excess. At fixed $\RH$ the factor $(1-\alpha\Ms/R^3)^{-1}$ in Eq.~(\ref{eq:S}) is largest near $R_0$ and decays as $R^{-2}$, so the holonomy correction adds a bounded amount that saturates for large black holes, whereas the bounce factor $(1-a^2/R^2)^{-1/2}$ produces the logarithm of Eq.~(\ref{eq:SSV}), which keeps growing as $\pi a^2\ln(2\RH/a)$ for $\RH\gg a$. The large-black-hole limit of the entropy is therefore $S\simeq\pi\RH^2+\pi a^2\ln(2\RH/a)+\pi\alpha\ln(\RH/R_0)+{\rm const}$, with both corrections logarithmic and of opposite geometric origin, one from the radial measure and one from the lapse. Logarithmic corrections of this form are familiar from quantum-corrected black hole entropies, and here they appear at the classical level of the effective geometry; since $\alpha$ is Planckian in the original derivation while $a$ is unconstrained, the bounce term is the one that could in principle be macroscopic. We also note that neither correction can be absorbed into a redefinition of the area, because the horizon area is $4\pi\RH^2$ in both coordinates.

{\setlength{\tabcolsep}{12pt}\renewcommand{\arraystretch}{1.6}\small
\begin{longtable}{c c c c c c c c}
\caption{Thermodynamic quantities at fixed $\alpha=0.5M^2$ as functions of the bounce parameter and the areal horizon radius. The entropy $S$ follows from the first law, and is compared with the area law $\pi R_h^2$; $C$ is the heat capacity and $F=M-T_HS$ the free energy.}\label{tab:thermo}\\
\cellcolor{brown!30}$a/M$ & \cellcolor{brown!30}$R_h/M$ & \cellcolor{brown!30}$M$ & \cellcolor{brown!30}$T_H M$ & \cellcolor{brown!30}$S/M^2$ & \cellcolor{brown!30}$\pi R_h^2/M^2$ & \cellcolor{brown!30}$C/M^2$ & \cellcolor{brown!30}$F/M$ \\ \hline\hline
\endfirsthead
\cellcolor{brown!30}$a/M$ & \cellcolor{brown!30}$R_h/M$ & \cellcolor{brown!30}$M$ & \cellcolor{brown!30}$T_H M$ & \cellcolor{brown!30}$S/M^2$ & \cellcolor{brown!30}$\pi R_h^2/M^2$ & \cellcolor{brown!30}$C/M^2$ & \cellcolor{brown!30}$F/M$ \\ \hline\hline
\endhead
\hline\hline
\endfoot
0.00 & 2.00 & 1.0334 & 0.03581 & 12.515 & 12.566 & -35.25 & 0.5853 \\
0.00 & 2.50 & 1.2761 & 0.02984 & 19.963 & 19.635 & -47.56 & 0.6803 \\
0.00 & 3.00 & 1.5214 & 0.02539 & 28.904 & 28.274 & -64.10 & 0.7876 \\
0.00 & 4.00 & 2.0159 & 0.01942 & 51.362 & 50.265 & -107.47 & 1.0184 \\
0.50 & 2.00 & 1.0334 & 0.03467 & 13.580 & 12.566 & -39.90 & 0.5626 \\
0.50 & 2.50 & 1.2761 & 0.02924 & 21.221 & 19.635 & -51.00 & 0.6556 \\
0.50 & 3.00 & 1.5214 & 0.02503 & 30.314 & 28.274 & -67.12 & 0.7626 \\
0.50 & 4.00 & 2.0159 & 0.01927 & 53.006 & 50.265 & -110.16 & 0.9945 \\
1.00 & 2.00 & 1.0334 & 0.03101 & 17.853 & 12.566 & -72.35 & 0.4798 \\
1.00 & 2.50 & 1.2761 & 0.02735 & 26.176 & 19.635 & -66.63 & 0.5602 \\
1.00 & 3.00 & 1.5214 & 0.02394 & 35.777 & 28.274 & -78.84 & 0.6650 \\
1.00 & 4.00 & 2.0159 & 0.01880 & 59.222 & 50.265 & -119.36 & 0.9023 \\
1.50 & 2.00 & 1.0334 & 0.02368 & 15.390 & 12.566 & 77.58 & 0.6689 \\
1.50 & 2.50 & 1.2761 & 0.02387 & 25.445 & 19.635 & -171.49 & 0.6686 \\
1.50 & 3.00 & 1.5214 & 0.02199 & 36.128 & 28.274 & -116.95 & 0.7271 \\
1.50 & 4.00 & 2.0159 & 0.01800 & 61.019 & 50.265 & -140.04 & 0.9173 \\
1.90 & 2.00 & 1.0334 & 0.01118 & 8.152 & 12.566 & 10.13 & 0.9422 \\
1.90 & 2.50 & 1.2761 & 0.01939 & 22.888 & 19.635 & 124.37 & 0.8322 \\
1.90 & 3.00 & 1.5214 & 0.01965 & 35.319 & 28.274 & -315.83 & 0.8275 \\
1.90 & 4.00 & 2.0159 & 0.01709 & 62.153 & 50.265 & -176.11 & 0.9537 \\
\end{longtable}}

The two parameters can also be promoted to thermodynamic variables. Under the rescaling $M\to\lambda M$, $\RH\to\lambda\RH$, $\alpha\to\lambda^2\alpha$, $a\to\lambda a$ the mass function~(\ref{eq:MofR}) is homogeneous of degree one and the entropy~(\ref{eq:S}) of degree two, so Euler's theorem applied to $M(S,\alpha,a)$ gives the Smarr-type relation
\begin{equation}
M=2T_HS+2\alpha\,\Phi_\alpha+a\,\Phi_a,\qquad \Phi_\alpha=\Big(\frac{\partial M}{\partial\alpha}\Big)_{S,a},\quad \Phi_a=\Big(\frac{\partial M}{\partial a}\Big)_{S,\alpha},
\label{eq:smarr}
\end{equation}
together with the extended first law $\dd M=T_H\dd S+\Phi_\alpha\dd\alpha+\Phi_a\dd a$. The two potentials follow from the chain rule at fixed $\RH$: $\Phi_\alpha=(\partial_\alpha M)_{\RH}-T_H(\partial_\alpha S)_{\RH}$ and $\Phi_a=-T_H(\partial_aS)_{\RH}$, since $M$ does not depend on $a$ at fixed areal horizon. The sign of $\Phi_a$ is not fixed. For $\alpha=0$ the entropy~(\ref{eq:SSV}) gives $(\partial_aS)_{\RH}=2\pi a\big[\ln\tfrac{\RH+\sqrt{\RH^2-a^2}}{a}-\tfrac{\RH}{\sqrt{\RH^2-a^2}}\big]$, which is positive for small $a$, where the logarithm dominates, and negative as $a\to\RH$, where the second term diverges; hence $\Phi_a=-T_H(\partial_aS)_{\RH}$ is negative for small bounces and positive near the throat-horizon coincidence, vanishing in between, and we verified symbolically that Eq.~(\ref{eq:smarr}) then reduces to $M=\RH/2$ identically. The situation is the same as for the regularization scale of Bardeen and Hayward black holes~\cite{Ma:2014,Simovic:2024}, whose conjugate potentials also depend on the branch. We do not pursue the extended phase space further here, since without a cosmological constant there is no pressure term and no van der Waals-type criticality to look for.

The topological classification of Wei, Liu and Mann~\cite{Wei:2022} treats every black hole branch as a defect of a vector field built from the generalized free energy $\mathcal{F}=\Ms-S/\tau$, where $\tau$ is an arbitrary inverse temperature. The field is $\phi=(\partial_{\RH}\mathcal{F},\,-\cot\Theta\csc\Theta)$ on the half-plane $\RH>R_{\rm min}$, $0<\Theta<\pi$, with the auxiliary angle introduced so that $\phi$ points outward at $\Theta=0,\pi$. Using the first law $\partial_{\RH}S=(\dd\Ms/\dd\RH)/T_H$ the radial component is
\begin{equation}
\phi^{R}=\frac{\dd\Ms}{\dd\RH}\left(1-\frac{1}{T_H\,\tau}\right),
\label{eq:phiR}
\end{equation}
so the zero points at $\Theta=\pi/2$ are the solutions of $\tau=1/T_H(\RH)$. Each zero carries a winding number $w=\pm1$ given by the sign of $\dd T_H/\dd\RH$, positive for locally stable branches ($C>0$) and negative for unstable ones ($C<0$), and the topological charge $W$ is the sum over all zeros; it is invariant under deformations of $\tau$ and characterizes the family rather than a single solution~\cite{Wei:2022,Duan:1979}. For Schwarzschild $\tau=4\pi\RH$ is monotonic, there is one zero with $w=-1$, and $W=-1$; for Reissner-Nordstr\"om and for most regular black holes $\tau(\RH)$ has a minimum, there are two zeros with opposite signs, and $W=0$~\cite{Wei:2022,Chen:2026,Xia:2026}.

\begin{figure}[!htbp]
\centering
\includegraphics[width=\textwidth]{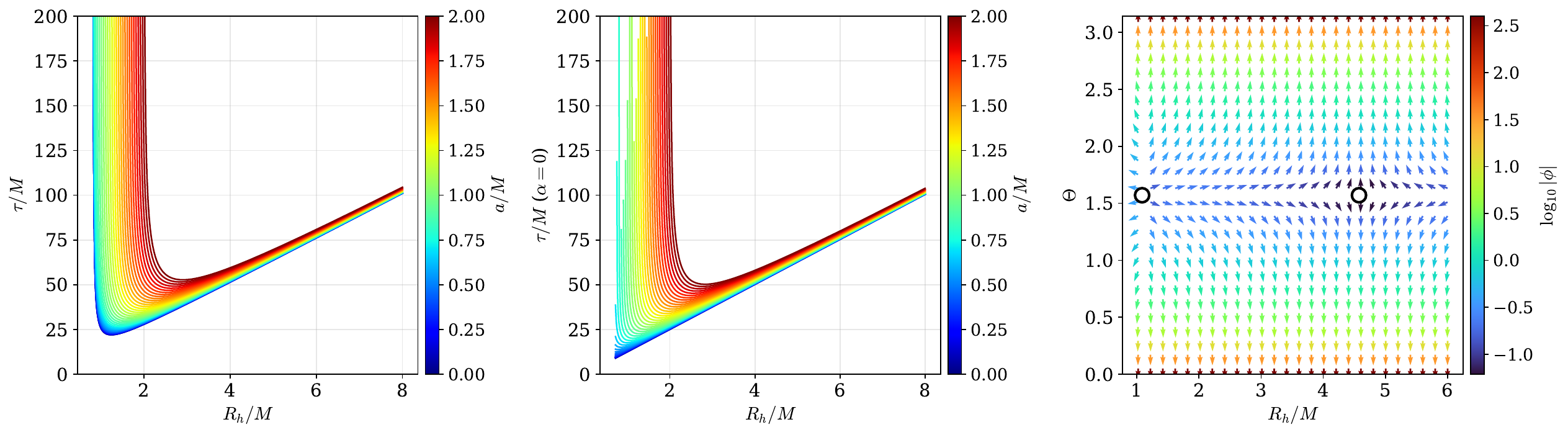}
\caption{Left: the defect curves $\tau=1/T_H$ at $\alpha=0.5M^2$ for $a$ from $0$ to $2M$; every curve diverges at $R_{\rm min}$ and grows linearly at large $R_h$, so a horizontal line $\tau={\rm const}$ above the minimum cuts each curve twice. Middle: the same curves for the Simpson-Visser family ($\alpha=0$), where the minimum is at $R_h=\sqrt2\,a$ and $\tau_{\min}=8\pi a$. Right: the unit vector field $\phi/|\phi|$ of Eq.~(\ref{eq:phiR}) at the reference point with $\tau=60M$, colored by $\log_{10}|\phi|$; the two zero points (circles) at $R_h=1.084M$ and $R_h=4.575M$ carry winding numbers $+1$ and $-1$.}
\label{fig:topo}
\end{figure}

Figure~\ref{fig:topo} shows the defect curves for the bounce-qOSD family and, separately, for its Simpson-Visser limit. In both cases $\tau(\RH)$ diverges at $R_{\rm min}$, because $T_H\to0$ there, and grows linearly at large $\RH$, so it has a single minimum and every horizontal line above it intersects the curve twice. The zero at smaller $\RH$ lies on the branch with $C>0$ and carries $w=+1$; the one at larger $\RH$ lies on the branch with $C<0$ and carries $w=-1$. Hence $W=0$ for every $a>0$ and every $0\leq\alpha<\alpha_{\rm ext}$. Two remarks follow. First, at $a=0$ and $\alpha>0$ the qOSD black hole already has $W=0$ because of its inner horizon, in agreement with the Reissner-Nordstr\"om-like classification of regular black holes~\cite{Wei:2022}, so the bounce does not change the charge of the seed geometry. Second, at $\alpha=0$ the seed is Schwarzschild with $W=-1$, and switching on any $a>0$ changes the charge to $W=0$. The Simpson-Visser deformation is therefore a topological transmutation in the thermodynamic sense, driven by the zero of $T_H$ at $\RH=a$ rather than by an inner horizon. For $\alpha=0$ the minimum of $\tau$ can be located analytically: $T_H\propto\sqrt{1-a^2/\RH^2}/\RH$ is maximal at $\RH=\sqrt2\,a$, where $T_H=1/(8\pi a)$, so that $\tau_{\min}=8\pi a$; our numerical values, $\RH=1.414M$ and $\tau_{\min}=25.13M$ at $a=M$, reproduce this. At the reference point the minimum sits at $\RH=1.638M$ with $\tau_{\min}=30.74M$, and the right panel of Fig.~\ref{fig:topo} displays the vector field with the two zero points for $\tau=60M$. We stress that the charge is a property of the black-hole sector only; the wormholes of Fig.~\ref{fig:phase} carry no horizon and no temperature and are outside the classification.

\section{Null geodesics, photon sphere, shadow and Lyapunov exponent}\label{isec4}

We now analyze photon motion, and we begin with the observation that organizes the rest of the paper. For the metric~(\ref{eq:metric}) with $h(r)=r^2+a^2$, equatorial null geodesics obey $\dot r^2=E^2-AL^2/h$, and circular orbits require $(A/h)'=0$. Since $\dd/\dd r=(r/R)\,\dd/\dd R$ and $r/R\neq0$ away from the throat, the condition is $\dd(A/R^2)/\dd R=0$, which involves the areal lapse only. The photon sphere therefore sits at the same areal radius as in the qOSD metric, given by the real root larger than $R_+$ of
\begin{equation}
\Rps^4-3\Ms\Rps^3+3\alpha\Ms^2=0,
\label{eq:ps}
\end{equation}
which reduces to $\Rps=3\Ms$ at $\alpha=0$ and decreases monotonically with $\alpha$ down to $\Rps=2.759\Ms$ at $\alpha_{\rm ext}$. The critical impact parameter and the coordinate angular velocity of the orbit are likewise $a$-independent,
\begin{equation}
b_c=\frac{\Rps}{\sqrt{A(\Rps)}},\qquad \Omega_c=\frac{\sqrt{A(\Rps)}}{\Rps},
\label{eq:bc}
\end{equation}
and since $A\to1$ at infinity the shadow radius seen by a distant observer is $r_{\rm sh}=b_c$~\cite{Perlick:2022,Cunha:2020}. The bounce coordinate of the photon sphere, $\rps=\sqrt{\Rps^2-a^2}$, does move with $a$, and the photon sphere exists as long as $a<\Rps$; for $\Rps<a$ the throat itself becomes the innermost circular null orbit, a familiar feature of black-bounce wormholes~\cite{Tsukamoto:2021a,Guerrero:2021}, which we do not pursue here.

The Lyapunov exponent that governs the instability of the null circular orbit, and through the eikonal correspondence the damping of high-multipole quasinormal modes~\cite{Cardoso:2009}, is where the bounce enters. In the coordinate $r$ the exponent is $\lambda_L^2=-\tfrac12\,A\,h\,(A/h)''$ evaluated at $\rps$, and converting the second derivative to the areal radius at a point where the first derivative vanishes gives $(A/h)''_{rr}=(1-a^2/R^2)(A/R^2)''_{RR}$. Hence
\begin{equation}
\lambda_L(\alpha,a)=\sqrt{1-\frac{a^2}{\Rps^2}}\;\lambda_L^{\rm qOSD}(\alpha),\qquad
\lambda_L^{\rm qOSD}=\sqrt{\frac{A(\Rps)}{2\Rps^2}\Big[2A(\Rps)-\Rps^2A''(\Rps)\Big]},
\label{eq:lyap}
\end{equation}
where the qOSD expression is the standard one of~\cite{Cardoso:2009}. We verified numerically that Eq.~(\ref{eq:lyap}) is exact, to twelve digits, against a direct finite-difference evaluation of the curvature of $A/h$ in $r$. At $\alpha=0$ it gives $\lambda_L=\sqrt{1-a^2/9\Ms^2}/(3\sqrt3\Ms)$ for the Simpson-Visser spacetime, and at $a=0$ the Schwarzschild value $1/(3\sqrt3\Ms)$. The decoupling is now explicit: the shadow and the real part of the eikonal frequency depend on $\alpha$ only, the imaginary part on both, and the ratio ${\rm Im}\,\omega/{\rm Re}\,\omega$ carries a clean $a$ signature.

\begin{figure}[!htbp]
\centering
\includegraphics[width=\textwidth]{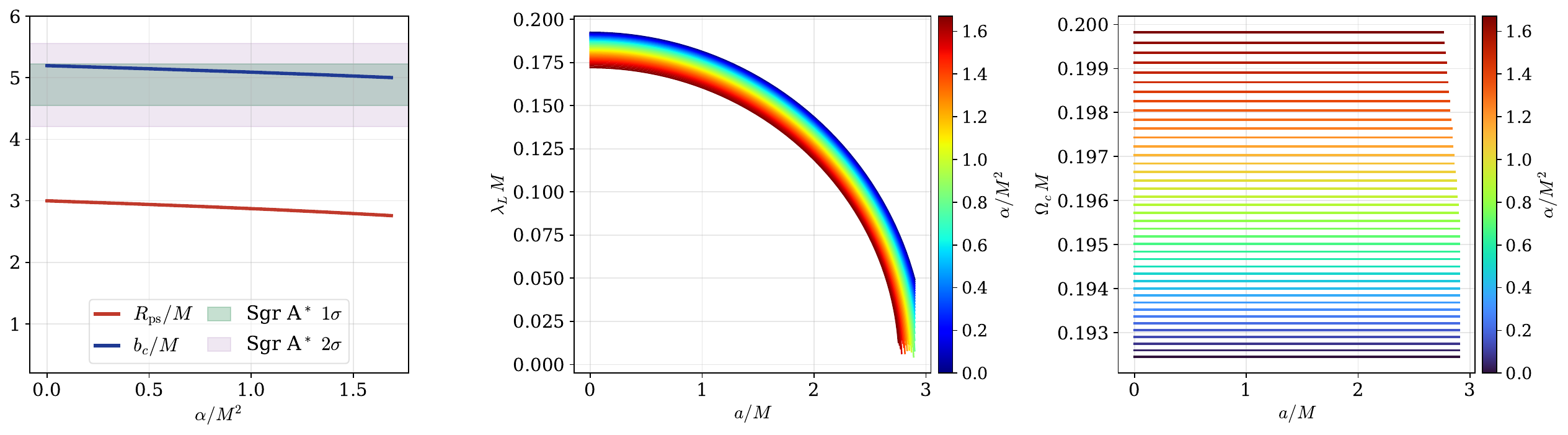}
\caption{Left: photon-sphere radius and critical impact parameter as functions of $\alpha$, with the $1\sigma$ and $2\sigma$ Sgr~A$^*$ bands for $r_{\rm sh}/M$ shaded; both quantities are independent of $a$. Middle: Lyapunov exponent as a function of $a$ for $\alpha$ from $0$ to $\alpha_{\rm ext}$, terminating where $a=R_{\rm ps}$. Right: orbital frequency $\Omega_c$ of the photon sphere on the same grid, which is flat in $a$.}
\label{fig:shadow}
\end{figure}

Figure~\ref{fig:shadow} and Tables~\ref{tab:geo} and~\ref{tab:lam} give the numbers. The impact parameter decreases from $b_c=3\sqrt3\Ms=5.196\Ms$ at $\alpha=0$ to $5.002\Ms$ at $\alpha\to\alpha_{\rm ext}$, a change of $3.7\%$ over the whole physical range. The Sgr~A$^*$ shadow, combined with the Keck and VLTI mass-to-distance priors, restricts the shadow radius to $4.55\lesssim r_{\rm sh}/\Ms\lesssim5.22$ at $1\sigma$ and $4.21\lesssim r_{\rm sh}/\Ms\lesssim5.56$ at $2\sigma$~\cite{Akiyama:2022,Vagnozzi:2023}, so the whole family is compatible with the observation and the shadow places no bound on $\alpha$ and none at all on $a$. This agrees with the qOSD analyses of~\cite{Yang:2023,Luo:2024,Yang:2026} and shows that the shadow alone cannot separate the bounce-regularized geometry from Schwarzschild; the discriminating power lies in the time-dependent and orbital observables of the next sections. The Lyapunov exponent, on the other hand, falls from $0.1884/\Ms$ at $a=0$ to $0.0992/\Ms$ at $a=2.5\Ms$ for $\alpha=0.5\Ms^2$, and vanishes as $a\to\Rps$. The suppression factor in Table~\ref{tab:lam} is $0.940$ at the reference point, meaning a ringdown that lasts $6\%$ longer than the qOSD one at the same $\alpha$ and $9\%$ longer than Schwarzschild.

{\setlength{\tabcolsep}{12pt}\renewcommand{\arraystretch}{1.6}\small
\begin{longtable}{c c c c c c c c}
\caption{Areal quantities that depend on $\alpha$ only: the two horizon radii, the photon-sphere radius $R_{\rm ps}$, the critical impact parameter $b_c$, the orbital frequency $\Omega_c$ of the null circular orbit, the Lyapunov exponent at $a=0$, and the ISCO radius. The last row is at $\alpha=0.999\,\alpha_{\rm ext}$.}\label{tab:geo}\\
\cellcolor{brown!30}$\alpha/M^2$ & \cellcolor{brown!30}$R_+/M$ & \cellcolor{brown!30}$R_-/M$ & \cellcolor{brown!30}$R_{\rm ps}/M$ & \cellcolor{brown!30}$b_c/M$ & \cellcolor{brown!30}$\Omega_c M$ & \cellcolor{brown!30}$\lambda_L M$ & \cellcolor{brown!30}$R_{\rm ISCO}/M$ \\ \hline\hline
\endfirsthead
\cellcolor{brown!30}$\alpha/M^2$ & \cellcolor{brown!30}$R_+/M$ & \cellcolor{brown!30}$R_-/M$ & \cellcolor{brown!30}$R_{\rm ps}/M$ & \cellcolor{brown!30}$b_c/M$ & \cellcolor{brown!30}$\Omega_c M$ & \cellcolor{brown!30}$\lambda_L M$ & \cellcolor{brown!30}$R_{\rm ISCO}/M$ \\ \hline\hline
\endhead
\hline\hline
\endfoot
0.000 & 2.0000 & -- & 3.0000 & 5.1962 & 0.1925 & 0.1925 & 6.0000 \\
0.250 & 1.9672 & 0.5575 & 2.9714 & 5.1716 & 0.1934 & 0.1906 & 5.9613 \\
0.500 & 1.9305 & 0.7336 & 2.9410 & 5.1459 & 0.1943 & 0.1884 & 5.9214 \\
0.750 & 1.8887 & 0.8731 & 2.9086 & 5.1189 & 0.1954 & 0.1859 & 5.8802 \\
1.000 & 1.8393 & 1.0000 & 2.8736 & 5.0904 & 0.1964 & 0.1830 & 5.8377 \\
1.250 & 1.7774 & 1.1272 & 2.8355 & 5.0602 & 0.1976 & 0.1796 & 5.7937 \\
1.500 & 1.6883 & 1.2732 & 2.7936 & 5.0280 & 0.1989 & 0.1755 & 5.7481 \\
1.686 & 1.5192 & 1.4805 & 2.7593 & 5.0024 & 0.1999 & 0.1718 & 5.7130 \\
\end{longtable}}

{\setlength{\tabcolsep}{12pt}\renewcommand{\arraystretch}{1.6}\small
\begin{longtable}{c c c c c c c c}
\caption{Effect of the bounce parameter at fixed $\alpha=0.5M^2$. The photon-sphere areal radius and the impact parameter are unchanged, whereas the Lyapunov exponent is suppressed by the factor $\sqrt{1-a^2/R_{\rm ps}^2}$ listed in the seventh column. The eighth column is the ISCO position in the bounce coordinate.}\label{tab:lam}\\
\cellcolor{brown!30}$a/M$ & \cellcolor{brown!30}$r_h/M$ & \cellcolor{brown!30}$r_{\rm ps}/M$ & \cellcolor{brown!30}$R_{\rm ps}/M$ & \cellcolor{brown!30}$b_c/M$ & \cellcolor{brown!30}$\lambda_L M$ & \cellcolor{brown!30}$\sqrt{1-a^2/R_{\rm ps}^2}$ & \cellcolor{brown!30}$r_{\rm ISCO}/M$ \\ \hline\hline
\endfirsthead
\cellcolor{brown!30}$a/M$ & \cellcolor{brown!30}$r_h/M$ & \cellcolor{brown!30}$r_{\rm ps}/M$ & \cellcolor{brown!30}$R_{\rm ps}/M$ & \cellcolor{brown!30}$b_c/M$ & \cellcolor{brown!30}$\lambda_L M$ & \cellcolor{brown!30}$\sqrt{1-a^2/R_{\rm ps}^2}$ & \cellcolor{brown!30}$r_{\rm ISCO}/M$ \\ \hline\hline
\endhead
\hline\hline
\endfoot
0.00 & 1.9305 & 2.9410 & 2.9410 & 5.1459 & 0.1884 & 1.0000 & 5.9214 \\
0.50 & 1.8646 & 2.8982 & 2.9410 & 5.1459 & 0.1857 & 0.9854 & 5.9002 \\
1.00 & 1.6513 & 2.7658 & 2.9410 & 5.1459 & 0.1772 & 0.9404 & 5.8363 \\
1.50 & 1.2153 & 2.5298 & 2.9410 & 5.1459 & 0.1621 & 0.8602 & 5.7282 \\
2.00 & -- & 2.1563 & 2.9410 & 5.1459 & 0.1381 & 0.7332 & 5.5734 \\
2.50 & -- & 1.5491 & 2.9410 & 5.1459 & 0.0992 & 0.5267 & 5.3677 \\
\end{longtable}}

\begin{figure}[!htbp]
\centering
\includegraphics[width=0.9\textwidth]{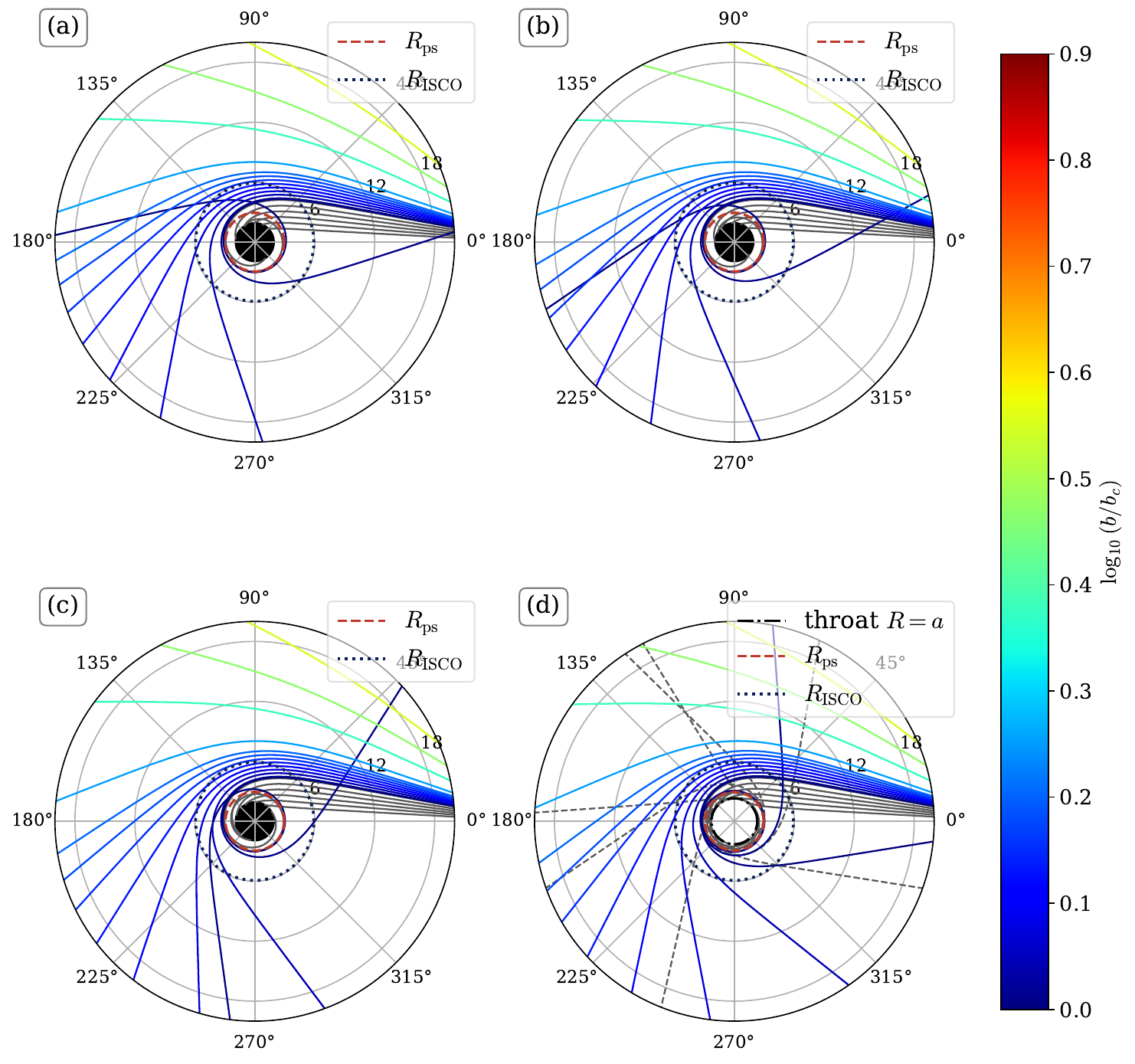}
\caption{Null geodesics at $\alpha=0.5M^2$ in the areal radius, for (a) the qOSD seed $a=0$, (b) the reference black hole $a=M$, (c) $a=1.8M$, close to the horizon-throat coincidence, and (d) the wormhole $a=2.3M$. Rays with $b>b_c$ are colored by $\log_{10}(b/b_c)$; rays with $b<b_c$ are gray and are captured. The black disk is the horizon $R_+$, the dashed circle the photon sphere and the dotted circle the ISCO; in panel (d) the dash-dotted circle is the throat and the dashed continuations of the captured rays lie on the second sheet $r<0$.}
\label{fig:rays}
\end{figure}

It is instructive to see how the rays themselves behave across the causal types, since the numbers above summarize only the critical orbit. We integrated the null geodesic equations $\ddot r=-\tfrac{b^2}{2}\,\dd(A/h)/\dd r$ and $\dot\phi=b/h$ for a fan of impact parameters launched from $r=28M$, and Fig.~\ref{fig:rays} draws the trajectories in the areal radius $R$ for the qOSD seed, for two black holes of the hidden-bounce type and for a wormhole, all at $\alpha=0.5M^2$. Rays with $b>b_c$ are colored by $\log_{10}(b/b_c)$ and turn around at the radius where $A/R^2=1/b^2$; those with $b<b_c$ are drawn in gray and cross the photon sphere. In the three black-hole panels the captured rays end on the horizon, and since $b_c$ and $R_{\rm ps}$ are the same in all three, the outer parts of the fans are identical; the only difference is the size of the horizon disk, which shrinks in $r$ but not in $R$, so the panels look alike outside $R_+$ by construction. This is the visual counterpart of the shadow degeneracy of Sec.~\ref{isec4}. The wormhole panel is different. There is no horizon, the throat is the dash-dotted circle $R=a$, and the captured rays reach it with finite $\dot r$ and continue into the second asymptotic region, drawn dashed; a distant observer on the first sheet sees them disappear, so the wormhole casts the same shadow as the black holes, but light from sources on the other sheet can reach the observer through the throat, which is the feature exploited in images of black-bounce wormholes illuminated from both sides~\cite{Guerrero:2021,Tsukamoto:2021a}. The critical rays at $b=b_c(1\pm10^{-3})$ wind several times around the photon sphere before escaping or falling in, and the number of windings per unit $\ln|b-b_c|$ is set by the Lyapunov exponent, which is why the bounce, invisible in the shadow, is visible in the photon-ring structure through Eq.~(\ref{eq:lyap}).

\section{Circular timelike orbits, ISCO and epicyclic frequencies}\label{isec5}

Massive particles on equatorial orbits obey $\dot r^2=E^2-V_{\rm eff}$ with $V_{\rm eff}=A(1+L^2/h)$, and the same argument as for null rays shows that the circular-orbit conditions $V_{\rm eff}=E^2$ and $\dd V_{\rm eff}/\dd R=0$ depend on $A(R)$ only. The specific energy and angular momentum of the circular orbit at areal radius $R$ are
\begin{equation}
E^2=\frac{2A^2}{2A-RA'},\qquad L^2=\frac{R^3A'}{2A-RA'},
\label{eq:EL}
\end{equation}
with primes now denoting $R$ derivatives, and the ISCO follows from $\dd L^2/\dd R=0$ or equivalently $\dd E^2/\dd R=0$. Both are $a$-independent, so the ISCO areal radius, energy and angular momentum are those of the qOSD black hole; the Schwarzschild values $R_{\rm ISCO}=6\Ms$, $E=2\sqrt2/3$, $L=2\sqrt3\Ms$ are recovered at $\alpha=0$, the procedure being the same as in our study of photon spheres and ISCOs around effective quantum-gravity black holes~\cite{Ahmed:2026d}, and at $\alpha=0.5\Ms^2$ we find $R_{\rm ISCO}=5.921\Ms$, $E_{\rm ISCO}=0.9422$, $L_{\rm ISCO}=3.452\Ms$. The Keplerian frequency measured by a distant observer, $\Omega_\phi^2=A'(R)/(2R)$, is also unchanged, and so is the vertical epicyclic frequency, which coincides with $\Omega_\phi$ for any spherically symmetric geometry. The radial epicyclic frequency, obtained from the second derivative of $V_{\rm eff}$ along the proper radial direction, picks up the bounce factor,
\begin{equation}
\Omega_r^2=\left(1-\frac{a^2}{R^2}\right)\frac{A^2}{2E^2}\,\frac{\dd^2V_{\rm eff}}{\dd R^2}\Big|_{L=L(R)}
=\left(1-\frac{a^2}{R^2}\right)\Omega_r^{\rm qOSD}(R)^2 ,
\label{eq:omr}
\end{equation}
by the same mechanism as in Eq.~(\ref{eq:lyap}). We verified this against the closed form for the Simpson-Visser case, $\Omega_r^2=\Omega_\phi^2(1-6\Ms/R)(1-a^2/R^2)$, at $R=8\Ms$ and $a=0.5\Ms$ to ten digits, and against the Schwarzschild epicyclic relation at $a=0$. Because $\Omega_r$ retains its zero at the ISCO, the bounce leaves the ISCO in place while flattening the radial oscillation everywhere outside it; the effect is largest near the ISCO, where $a/R$ is largest, and drops as $R^{-2}$ outward.

\begin{figure}[!htbp]
\centering
\includegraphics[width=\textwidth]{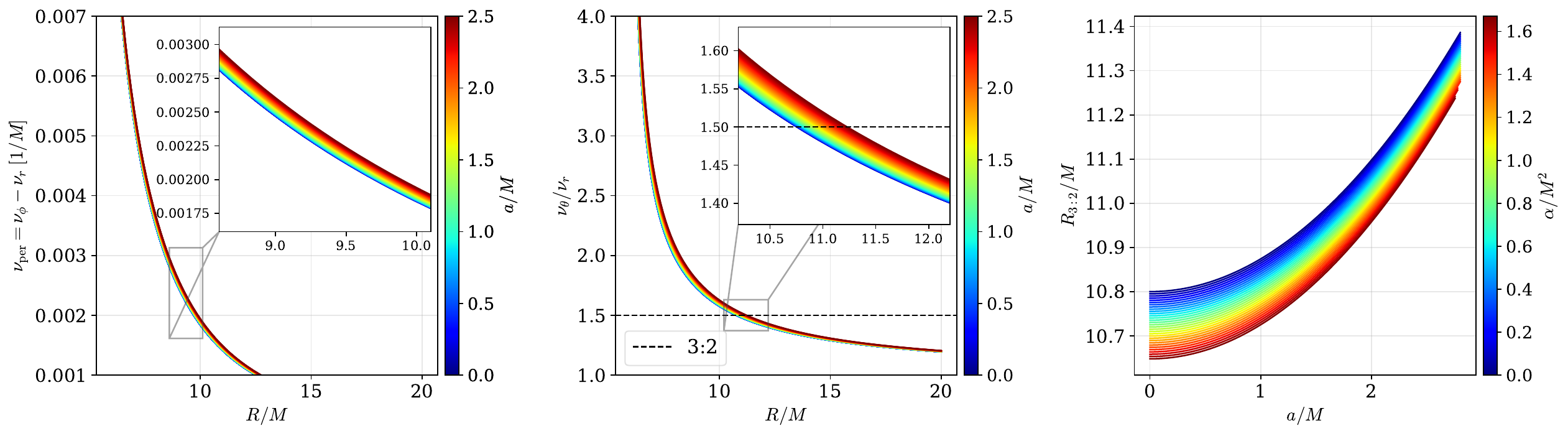}
\caption{Left: periastron precession frequency $\nu_{\rm per}=\nu_\phi-\nu_r$ at $\alpha=0.5M^2$ for $a$ from $0$ to $2.5M$, as a function of the areal radius; the bounce raises the precession because it lowers $\nu_r$. Middle: ratio $\nu_\theta/\nu_r$ on the same grid, with the $3{:}2$ line. The insets magnify the marked windows, by a factor of about eight in the left panel and six in the middle one, with their own axis labels, so that the ordering of the curves in $a$ becomes visible. Right: areal radius $R_{3:2}$ at which $\nu_\theta/\nu_r=3/2$, as a function of $a$ for $\alpha$ from $0$ to $\alpha_{\rm ext}$.}
\label{fig:epi}
\end{figure}

The relativistic precession model of high-frequency quasiperiodic oscillations~\cite{Stella:1998,Stella:1999} identifies the upper peak with $\nu_\phi=\Omega_\phi/2\pi$, the lower peak with the periastron precession frequency $\nu_{\rm per}=\nu_\phi-\nu_r$, and the low-frequency feature with the nodal precession $\nu_{\rm nod}=\nu_\phi-\nu_\theta$; the model has been used to extract the mass and spin of GRO~J1655-40 from simultaneous detections~\cite{Motta:2014} and has been applied to a range of modified and regular black holes~\cite{Stuchlik:2021,Bouzenada:2025,Mustafa:2026,Jumaniyozov:2025}. In the present geometry $\nu_{\rm nod}=0$ identically, so the model degenerates to two frequencies, and the bounce affects only $\nu_{\rm per}$ through $\nu_r$. Figure~\ref{fig:epi} shows that the periastron precession increases with $a$ at fixed $R$, that the ratio $\nu_\theta/\nu_r$ increases with $a$, and that consequently the radius at which the $3{:}2$ resonance of the twin-peak models~\cite{Torok:2005} occurs moves outward, from $R_{3:2}=10.755\Ms$ at $a=0$ to $10.83\Ms$ at the reference point and $11.23\Ms$ at $a=2.5\Ms$. The corresponding Schwarzschild value is $10.8\Ms$, and the $\alpha$ correction lowers it by about half a percent, so the two parameters push $R_{3:2}$ in opposite directions; Table~\ref{tab:epi} lists the numbers. Since the areal radius is what an accretion model measures through the Keplerian frequency, a joint determination of $\nu_\phi$ and $\nu_{\rm per}$ from a source with independently known mass would return $a$ directly through the factor in Eq.~(\ref{eq:omr}), with no degeneracy with $\alpha$ at leading order. The effect is small for astrophysical bounces, at the level of a percent in $\nu_r$ for $a\sim\Ms$ near the ISCO, but it is the only geodesic observable in this family that sees the bounce at all.

{\setlength{\tabcolsep}{12pt}\renewcommand{\arraystretch}{1.6}\small
\begin{longtable}{c c c c c c c c}
\caption{Circular timelike orbits at $\alpha=0.5M^2$: ISCO areal radius, specific energy and angular momentum at the ISCO, the azimuthal and radial epicyclic frequencies at $R=1.5R_{\rm ISCO}$, the analytic suppression factor of $\Omega_r$ at that radius, and the areal radius $R_{3:2}$ at which $\nu_\theta/\nu_r=3/2$.}\label{tab:epi}\\
\cellcolor{brown!30}$a/M$ & \cellcolor{brown!30}$R_{\rm ISCO}/M$ & \cellcolor{brown!30}$E_{\rm ISCO}$ & \cellcolor{brown!30}$L_{\rm ISCO}/M$ & \cellcolor{brown!30}$\Omega_\phi M$ & \cellcolor{brown!30}$\Omega_r M$ & \cellcolor{brown!30}$\sqrt{1-a^2/R^2}$ & \cellcolor{brown!30}$R_{3:2}/M$ \\ \hline\hline
\endfirsthead
\cellcolor{brown!30}$a/M$ & \cellcolor{brown!30}$R_{\rm ISCO}/M$ & \cellcolor{brown!30}$E_{\rm ISCO}$ & \cellcolor{brown!30}$L_{\rm ISCO}/M$ & \cellcolor{brown!30}$\Omega_\phi M$ & \cellcolor{brown!30}$\Omega_r M$ & \cellcolor{brown!30}$\sqrt{1-a^2/R^2}$ & \cellcolor{brown!30}$R_{3:2}/M$ \\ \hline\hline
\endhead
\hline\hline
\endfoot
0.00 & 5.9214 & 0.9422 & 3.4518 & 0.03775 & 0.02164 & 1.0000 & 10.7555 \\
0.50 & 5.9214 & 0.9422 & 3.4518 & 0.03775 & 0.02160 & 0.9984 & 10.7742 \\
1.00 & 5.9214 & 0.9422 & 3.4518 & 0.03775 & 0.02150 & 0.9936 & 10.8306 \\
1.50 & 5.9214 & 0.9422 & 3.4518 & 0.03775 & 0.02133 & 0.9856 & 10.9247 \\
2.00 & 5.9214 & 0.9422 & 3.4518 & 0.03775 & 0.02108 & 0.9743 & 11.0570 \\
2.50 & 5.9214 & 0.9422 & 3.4518 & 0.03775 & 0.02076 & 0.9596 & 11.2276 \\
\end{longtable}}
\section{Radial infall, tidal forces and passage through the throat}\label{isec5b}

We close the geodesic analysis by following an observer who falls radially through the horizon and asks what happens at the bounce, which is the place where the qOSD geometry and the regularized one differ most. Radial timelike geodesics obey $\dot r^2=E^2-A(r)$ with $E^2=A(r_0)$ for an observer released from rest at $r_0$, and the proper time between two radii is $\tau=\int\dd r/\sqrt{E^2-A}$. Since $A$ is bounded on the whole axis, the integrand is finite everywhere except at turning points, and the geodesic either reaches the throat and continues into the second asymptotic region, or turns around where $A(r)=E^2$. Which of the two happens is decided by the sign of $A$ at the throat. In the hidden-bounce region $A(a)<0<E^2$, so $\dot r^2>0$ at $r=0$ and the observer crosses the throat with a finite radial velocity; the throat is a spacelike surface in that region, crossed once, and the geodesic emerges into the $r<0$ copy of the exterior, where $A$ turns positive again at $r=-\rH$. This is the standard picture of the black bounce~\cite{Simpson:2019,Lobo:2021}, and it is the sense in which the geometry is geodesically complete; the analysis of~\cite{Zhou:2023} applies to regular black holes with a de Sitter core, not to bounces. In the two-horizon region $A(a)>0$ and the throat sits inside the inner horizon where $A$ grows steeply, so a geodesic with $E^2<A(a)$ has a turning point between the inner horizon and the throat and never reaches $r=0$, as in Reissner-Nordstr\"om; only observers with $E^2>A(a)$, which requires $E^2>1$ for the black holes in Table~\ref{tab:regimes} with $a<R_-$, cross the throat. In the wormhole region $0<A(a)<E^2$ for any observer released from rest outside the minimum of $A$, so the throat is crossed freely in both directions.

Tidal forces along the infall follow from the geodesic deviation equation in the comoving frame. For a radial geodesic with energy $E$ the deviation vector obeys $\ddot\xi^{\hat r}=\mathcal{K}_{\rm rad}\,\xi^{\hat r}$ and $\ddot\xi^{\hat\theta}=\mathcal{K}_{\rm ang}\,\xi^{\hat\theta}$, and in terms of the sectional curvatures of Sec.~\ref{isec2} one finds
\begin{equation}
\mathcal{K}_{\rm rad}=E_1=-\frac{A''}{2},\qquad
\mathcal{K}_{\rm ang}=E_2+\frac{E^2-A}{A}\,(E_2-E_3)=-\frac{A'R'}{2R}+\big(E^2-A\big)\frac{a^2}{R^4},
\label{eq:tidal}
\end{equation}
where the radial eigenvalue is boost invariant and the angular one is boosted by the Lorentz factor $\gamma^2v^2=(E^2-A)/A$ of the observer relative to the static frame. The combination $(E_2-E_3)/A=R''/R=a^2/R^4$ is regular at the horizon, so Eq.~(\ref{eq:tidal}) is finite across $A=0$ and the second form is the one we evaluate. For Schwarzschild $E_2=E_3$ and the familiar $\mathcal{K}_{\rm rad}=2M/r^3$, $\mathcal{K}_{\rm ang}=-M/r^3$ are recovered; at $\alpha=0$ Eq.~(\ref{eq:tidal}) agrees with the Simpson-Visser tidal tensor of~\cite{Arora:2024}. At the throat both eigenvalues reduce to closed expressions,
\begin{equation}
\mathcal{K}_{\rm rad}(0)=-\frac{1}{2a}\frac{\dd A}{\dd R}\Big|_{R=a}=-\frac{M}{a^3}\left(1-\frac{2\alpha M}{a^3}\right),\qquad
\mathcal{K}_{\rm ang}(0)=\frac{E^2-A(a)}{a^2},
\label{eq:tidal0}
\end{equation}
so the radial tidal eigenvalue at the throat changes sign at $a^3=2\alpha M$, being compressive for larger $a$ and stretching for smaller, whereas the angular one is always positive, that is, stretching, for an observer that actually reaches the throat. At the reference point $a^3=2\alpha M$ holds exactly and the radial tidal force vanishes at the throat, an accident of our parameter choice that Table~\ref{tab:infall} makes visible.

\begin{figure}[!htbp]
\centering
\includegraphics[width=\textwidth]{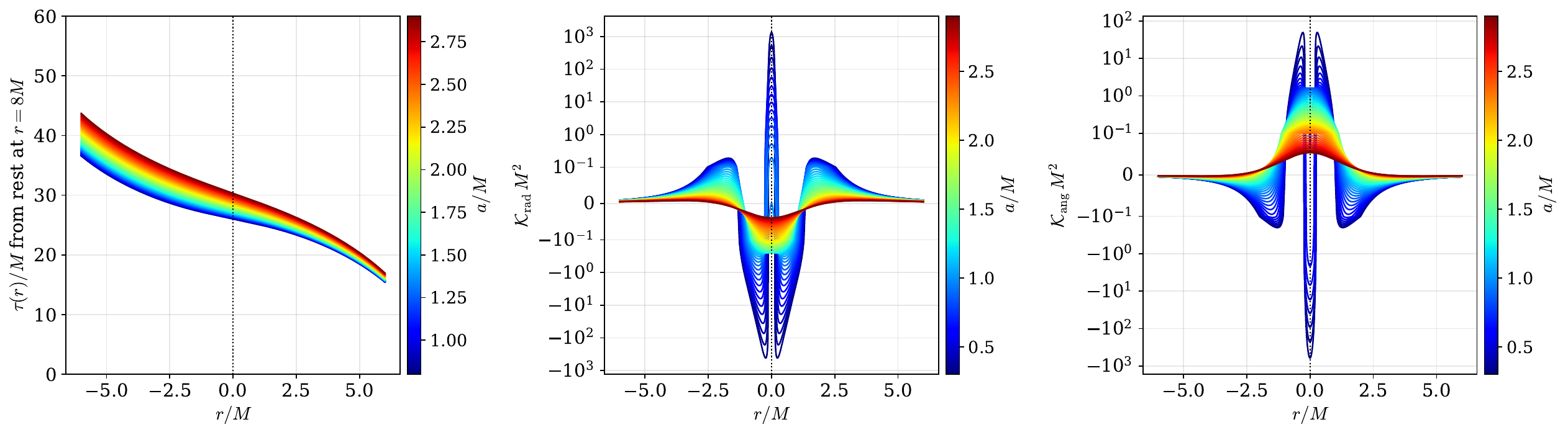}
\caption{Radial infall from rest at $r=8M$ at $\alpha=0.5M^2$. Left: proper time as a function of position for $a$ from $0.8M$ to $2.9M$, all of which cross the throat (dotted line) and continue into the second asymptotic region; larger $a$ means a longer trip because the gravitational pull is weaker at fixed $r$. Middle and right: radial and angular tidal eigenvalues of Eq.~(\ref{eq:tidal}) along the same trajectories for $a$ from $0.3M$ to $2.9M$, on symmetric logarithmic scales. Both are finite everywhere and largest near the throat for the smallest $a$, and both change sign with $a$ at the throat in accordance with Eq.~(\ref{eq:tidal0}).}
\label{fig:infall}
\end{figure}

{\setlength{\tabcolsep}{12pt}\renewcommand{\arraystretch}{1.6}\small
\begin{longtable}{c c c c c c}
\caption{Radial infall from rest at $r=8M$ for $\alpha=0.5M^2$: proper time to reach the horizon and the throat, and the radial and angular tidal eigenvalues at the throat. For $a=0.5M$ the throat lies inside the inner horizon and is not reached, since the observer turns around at $A(r)=E^2$; for the wormholes ($a\geq2M$) there is no horizon and the two proper times coincide.}\label{tab:infall}\\
\cellcolor{brown!30}$a/M$ & \cellcolor{brown!30}$r_h/M$ & \cellcolor{brown!30}$\tau_{8M\to r_h}/M$ & \cellcolor{brown!30}$\tau_{8M\to 0}/M$ & \cellcolor{brown!30}$\mathcal{K}_{\rm rad}(0)M^2$ & \cellcolor{brown!30}$\mathcal{K}_{\rm ang}(0)M^2$ \\ \hline\hline
\endfirsthead
\cellcolor{brown!30}$a/M$ & \cellcolor{brown!30}$r_h/M$ & \cellcolor{brown!30}$\tau_{8M\to r_h}/M$ & \cellcolor{brown!30}$\tau_{8M\to 0}/M$ & \cellcolor{brown!30}$\mathcal{K}_{\rm rad}(0)M^2$ & \cellcolor{brown!30}$\mathcal{K}_{\rm ang}(0)M^2$ \\ \hline\hline
\endhead
\hline\hline
\endfoot
0.50 & 1.8646 & 24.006 & -- & 56.000 & -16.998 \\
1.00 & 1.6513 & 24.591 & 26.199 & -0.000 & 1.252 \\
1.50 & 1.2153 & 25.679 & 26.962 & -0.209 & 0.440 \\
2.00 & -- & 27.981 & 27.981 & -0.109 & 0.182 \\
2.50 & -- & 29.210 & 29.210 & -0.060 & 0.088 \\
\end{longtable}}

Figure~\ref{fig:infall} and Table~\ref{tab:infall} quantify the passage. An observer released from rest at $r=8M$ reaches the horizon after $\tau=24.6M$ at the reference point and the throat after $26.2M$, so the interval between horizon crossing and bounce is $1.6M$, comparable to the horizon-to-singularity time of $\sim 1.3M$ for a Schwarzschild hole of the same mass and starting radius; the difference is that here the observer survives, with tidal eigenvalues at the throat of order $M^{-2}$ rather than divergent. The tidal profile is symmetric in $r$ and peaks at the throat. For the smallest bounce in the figure, $a=0.3M$, the peak values reach $10^3/M^2$, which is still finite but large enough to disrupt any extended body and is the imprint of the $a^{-3}$ and $a^{-4}$ scalings in Eq.~(\ref{eq:tidal0}); for $a\gtrsim M$ they are of order $M^{-2}$, within an order of magnitude of the horizon value $1/(4M^2)$ of a Schwarzschild hole of the same mass, and for the wormholes they are milder still. We also note that the proper-time curves in the left panel are ordered monotonically in $a$ over the entire trajectory, which means that a clock carried through the bounce and compared with one left at large radius would measure $a$ directly; the effect is of order $a^2/r^2$ and only becomes appreciable inside a few gravitational radii.
\section{Perturbations and quasinormal spectra}\label{isec6}

We next study how test fields of spin $0$, $1$ and $1/2$ propagate on the geometry, and we obtain the quasinormal spectra with three independent methods. Throughout, the time dependence is $e^{-i\omega t}$, so that damped modes have ${\rm Im}\,\omega<0$ and a mode with ${\rm Im}\,\omega>0$ would signal an instability. The tortoise coordinate is $\dd r_*=\dd r/A$; for the black holes it maps $r\in(\rH,\infty)$ onto the real line, and for the wormholes it maps the whole axis $r\in(-\infty,\infty)$ onto the real line with the throat at a finite $r_*$.

A massless scalar field decomposed as $\Phi=\sum_{\ell m}\psi_\ell(r)Y_{\ell m}e^{-i\omega t}/R(r)$ obeys the Klein-Gordon equation $\Box\Phi=0$, which on the metric~(\ref{eq:metric}) reduces to the Schr\"odinger form $\dd^2\psi/\dd r_*^2+(\omega^2-V_s)\psi=0$ with
\begin{equation}
V_s=A\left[\frac{\ell(\ell+1)}{R^2}+\frac{1}{R}\frac{\dd^2R}{\dd r_*^2}\right]
=A\left[\frac{\ell(\ell+1)}{r^2+a^2}+\frac{r\,A'(r)}{r^2+a^2}+\frac{a^2A}{(r^2+a^2)^2}\right],
\label{eq:Vs}
\end{equation}
where we used $\dd R/\dd r_*=Ar/R$ and $A'=\dd A/\dd r$. The last term is the fingerprint of the bounce: it is absent at $a=0$, where Eq.~(\ref{eq:Vs}) becomes the usual $A[\ell(\ell+1)/r^2+A'/r]$, and it is the only piece of the potential that survives at the throat of a wormhole, since $A'(0)=0$. The electromagnetic field in the Regge-Wheeler gauge gives the simpler potential
\begin{equation}
V_{\rm em}=\frac{A\,\ell(\ell+1)}{r^2+a^2},
\label{eq:Vem}
\end{equation}
which follows from the conformal invariance of the Maxwell equations in four dimensions and involves no derivative of $R$. For a massless Dirac field the standard reduction in a spherically symmetric background~\cite{Cho:2003} yields the pair of supersymmetric partner potentials
\begin{equation}
V_\pm=W^2\pm\frac{\dd W}{\dd r_*},\qquad W=\frac{\sqrt{A}\,\kappa}{\sqrt{r^2+a^2}},\qquad \kappa=\ell+1=1,2,\dots,
\label{eq:Vd}
\end{equation}
which share the same quasinormal spectrum; we quote results for $V_+$. The two Dirac potentials are related by the supersymmetric structure $V_\pm=W^2\pm W_{,r_*}$, so any solution of one Schr\"odinger problem maps to a solution of the other through the operator $\dd/\dd r_*\pm W$, and the quasinormal frequencies coincide although the transmission amplitudes differ by a phase. We verified this numerically: the third-order WKB frequencies from $V_-$ agree with those from $V_+$ to the fourth digit, and the residual difference is a measure of the WKB error rather than a physical splitting. All three potentials vanish at the horizon and at infinity, are positive outside the horizon of every black-hole configuration in Fig.~\ref{fig:phase}, and satisfy the Schwarzschild limit; we checked that our third-order WKB implementation reproduces the Schwarzschild benchmarks $\omega\Ms=0.4832-0.0968i$ (scalar $\ell=2$), $0.4571-0.0951i$ (electromagnetic $\ell=2$) and $0.3786-0.0965i$ (Dirac $\kappa=2$) to the accuracy expected of the method~\cite{Iyer:1987,Konoplya:2019,Cho:2003}.

\begin{figure}[!htbp]
\centering
\includegraphics[width=\textwidth]{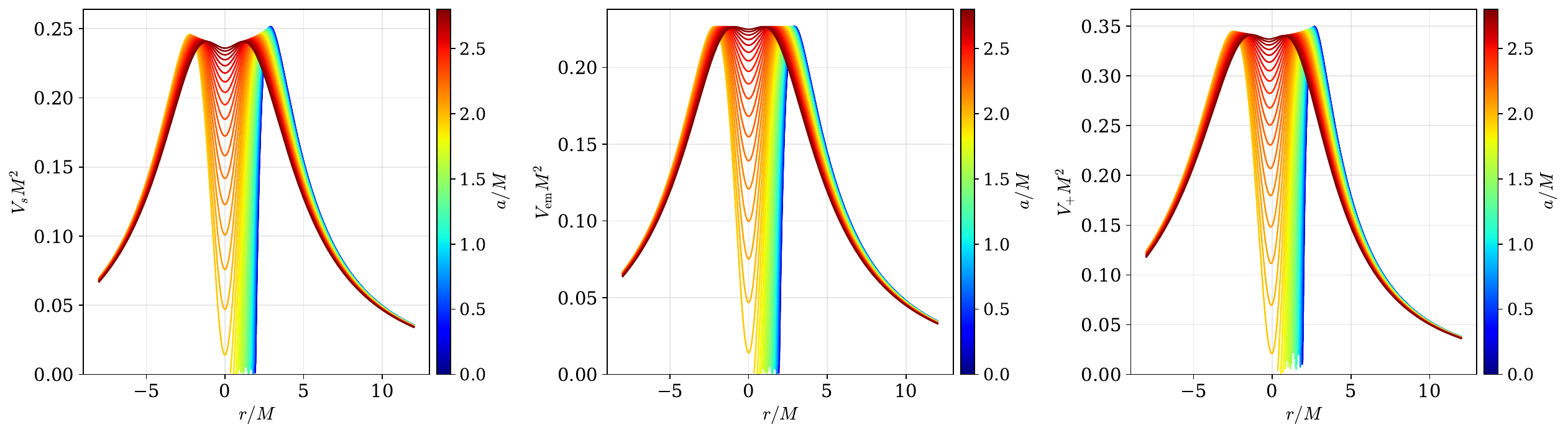}
\caption{Effective potentials for $\ell=2$ at $\alpha=0.5M^2$ as $a$ runs from $0$ to $2.8M$: scalar (left), electromagnetic (middle) and Dirac $V_+$ (right). For black-hole configurations ($a<R_+$) the curves are cut at the horizon and show a single barrier; for wormholes ($a>R_+$) the potential is defined on the whole axis, is symmetric in $r$, and develops two barriers separated by a well centered on the throat.}
\label{fig:pot}
\end{figure}

Figure~\ref{fig:pot} shows the three potentials across the black-hole-to-wormhole transition. In the black-hole regime the barrier sits near $\rps$ and its height is essentially fixed by $\alpha$, since it is dominated by the areal term $A\ell(\ell+1)/R^2$ evaluated at the photon sphere, whereas its width in $r_*$ depends on $a$. Once $a$ exceeds $R_+$ the horizon disappears, the potential becomes an even function of $r$ with two symmetric peaks at $\pm\rps$, and the region between them is a well whose depth is $A(a)\ell(\ell+1)/a^2$ plus the bounce term. Waves can be trapped in this well, which is the mechanism behind the echoes discussed below~\cite{Cardoso:2016,Churilova:2020,Bronnikov:2022,Yang:2023b}.

For the quasinormal frequencies we use the third-order WKB formula of Iyer and Will~\cite{Schutz:1985,Iyer:1987}, the same tool we applied to Dirac and scalar fields on Bardeen and dilaton backgrounds~\cite{AlBadawi:2021,Sakalli:2016,Sakalli:2015}, evaluated with the derivatives of the potential with respect to the tortoise coordinate computed symbolically through $\dd/\dd r_*=A\,\dd/\dd r$; higher orders~\cite{Konoplya:2003,Konoplya:2019} improve the accuracy for $\ell\geq2$ only marginally and we did not need them, while for $\ell=0$ the WKB method is known to be unreliable and we rely on the spectral solver instead. Tables~\ref{tab:qnms}, \ref{tab:qnmem} and~\ref{tab:qnmd} list the fundamental modes for $\ell=1,2,3$ and the first overtone of $\ell=2$ on a grid of $(\alpha,a)$ restricted to black-hole configurations.

{\setlength{\tabcolsep}{12pt}\renewcommand{\arraystretch}{1.6}\small
\begin{longtable}{c c c c c c}
\caption{Scalar quasinormal frequencies $\omega M$ from the third-order WKB formula. The first three frequency columns are the fundamental modes $n=0$ for $\ell=1,2,3$; the last column is the first overtone $n=1$ of $\ell=2$.}\label{tab:qnms}\\
\cellcolor{brown!30}$\alpha/M^2$ & \cellcolor{brown!30}$a/M$ & \cellcolor{brown!30}$\ell=1$ & \cellcolor{brown!30}$\ell=2$ & \cellcolor{brown!30}$\ell=3$ & \cellcolor{brown!30}$\ell=2,\,n=1$ \\ \hline\hline
\endfirsthead
\cellcolor{brown!30}$\alpha/M^2$ & \cellcolor{brown!30}$a/M$ & \cellcolor{brown!30}$\ell=1$ & \cellcolor{brown!30}$\ell=2$ & \cellcolor{brown!30}$\ell=3$ & \cellcolor{brown!30}$\ell=2,\,n=1$ \\ \hline\hline
\endhead
\hline\hline
\endfoot
0.00 & 0.00 & $0.2911-0.0980i$ & $0.4832-0.0968i$ & $0.6752-0.0965i$ & $0.4632-0.2958i$ \\
0.00 & 0.50 & $0.2910-0.0965i$ & $0.4832-0.0954i$ & $0.6752-0.0952i$ & $0.4639-0.2915i$ \\
0.00 & 1.00 & $0.2904-0.0920i$ & $0.4829-0.0912i$ & $0.6751-0.0910i$ & $0.4653-0.2781i$ \\
0.00 & 1.50 & $0.2890-0.0843i$ & $0.4823-0.0837i$ & $0.6746-0.0835i$ & $0.4653-0.2546i$ \\
0.50 & 0.00 & $0.2938-0.0954i$ & $0.4879-0.0946i$ & $0.6818-0.0944i$ & $0.4684-0.2888i$ \\
0.50 & 0.50 & $0.2936-0.0940i$ & $0.4878-0.0932i$ & $0.6818-0.0930i$ & $0.4690-0.2844i$ \\
0.50 & 1.00 & $0.2930-0.0895i$ & $0.4876-0.0889i$ & $0.6816-0.0888i$ & $0.4701-0.2709i$ \\
0.50 & 1.50 & $0.2914-0.0818i$ & $0.4868-0.0814i$ & $0.6811-0.0812i$ & $0.4694-0.2472i$ \\
1.00 & 0.00 & $0.2963-0.0921i$ & $0.4930-0.0917i$ & $0.6891-0.0916i$ & $0.4730-0.2796i$ \\
1.00 & 0.50 & $0.2961-0.0907i$ & $0.4929-0.0903i$ & $0.6890-0.0902i$ & $0.4735-0.2752i$ \\
1.00 & 1.00 & $0.2955-0.0863i$ & $0.4926-0.0860i$ & $0.6889-0.0859i$ & $0.4743-0.2616i$ \\
1.00 & 1.50 & $0.2938-0.0787i$ & $0.4917-0.0783i$ & $0.6883-0.0782i$ & $0.4729-0.2378i$ \\
1.50 & 0.00 & $0.2983-0.0880i$ & $0.4983-0.0878i$ & $0.6971-0.0878i$ & $0.4758-0.2677i$ \\
1.50 & 0.50 & $0.2982-0.0866i$ & $0.4982-0.0864i$ & $0.6971-0.0864i$ & $0.4762-0.2633i$ \\
1.50 & 1.00 & $0.2976-0.0823i$ & $0.4979-0.0821i$ & $0.6969-0.0820i$ & $0.4767-0.2497i$ \\
1.50 & 1.50 & $0.2959-0.0748i$ & $0.4970-0.0743i$ & $0.6962-0.0742i$ & $0.4745-0.2259i$ \\
\end{longtable}}

{\setlength{\tabcolsep}{12pt}\renewcommand{\arraystretch}{1.6}\small
\begin{longtable}{c c c c c c}
\caption{Electromagnetic quasinormal frequencies $\omega M$ (third-order WKB), same layout as Table~\ref{tab:qnms}.}\label{tab:qnmem}\\
\cellcolor{brown!30}$\alpha/M^2$ & \cellcolor{brown!30}$a/M$ & \cellcolor{brown!30}$\ell=1$ & \cellcolor{brown!30}$\ell=2$ & \cellcolor{brown!30}$\ell=3$ & \cellcolor{brown!30}$\ell=2,\,n=1$ \\ \hline\hline
\endfirsthead
\cellcolor{brown!30}$\alpha/M^2$ & \cellcolor{brown!30}$a/M$ & \cellcolor{brown!30}$\ell=1$ & \cellcolor{brown!30}$\ell=2$ & \cellcolor{brown!30}$\ell=3$ & \cellcolor{brown!30}$\ell=2,\,n=1$ \\ \hline\hline
\endhead
\hline\hline
\endfoot
0.00 & 0.00 & $0.2459-0.0931i$ & $0.4571-0.0951i$ & $0.6567-0.0956i$ & $0.4358-0.2910i$ \\
0.00 & 0.50 & $0.2465-0.0917i$ & $0.4575-0.0937i$ & $0.6570-0.0943i$ & $0.4370-0.2867i$ \\
0.00 & 1.00 & $0.2483-0.0872i$ & $0.4585-0.0895i$ & $0.6577-0.0901i$ & $0.4400-0.2733i$ \\
0.00 & 1.50 & $0.2507-0.0792i$ & $0.4598-0.0819i$ & $0.6586-0.0826i$ & $0.4426-0.2492i$ \\
0.50 & 0.00 & $0.2492-0.0905i$ & $0.4622-0.0929i$ & $0.6636-0.0935i$ & $0.4416-0.2839i$ \\
0.50 & 0.50 & $0.2499-0.0891i$ & $0.4625-0.0915i$ & $0.6638-0.0921i$ & $0.4427-0.2796i$ \\
0.50 & 1.00 & $0.2517-0.0847i$ & $0.4635-0.0872i$ & $0.6645-0.0879i$ & $0.4454-0.2660i$ \\
0.50 & 1.50 & $0.2540-0.0765i$ & $0.4648-0.0795i$ & $0.6654-0.0802i$ & $0.4473-0.2416i$ \\
1.00 & 0.00 & $0.2527-0.0870i$ & $0.4677-0.0899i$ & $0.6711-0.0907i$ & $0.4469-0.2745i$ \\
1.00 & 0.50 & $0.2533-0.0856i$ & $0.4680-0.0885i$ & $0.6714-0.0893i$ & $0.4479-0.2701i$ \\
1.00 & 1.00 & $0.2551-0.0812i$ & $0.4690-0.0842i$ & $0.6720-0.0850i$ & $0.4503-0.2564i$ \\
1.00 & 1.50 & $0.2574-0.0730i$ & $0.4702-0.0763i$ & $0.6729-0.0772i$ & $0.4514-0.2317i$ \\
1.50 & 0.00 & $0.2557-0.0819i$ & $0.4736-0.0857i$ & $0.6795-0.0867i$ & $0.4502-0.2617i$ \\
1.50 & 0.50 & $0.2564-0.0806i$ & $0.4739-0.0843i$ & $0.6798-0.0853i$ & $0.4511-0.2573i$ \\
1.50 & 1.00 & $0.2583-0.0763i$ & $0.4749-0.0800i$ & $0.6805-0.0809i$ & $0.4532-0.2435i$ \\
1.50 & 1.50 & $0.2606-0.0682i$ & $0.4761-0.0721i$ & $0.6813-0.0730i$ & $0.4536-0.2190i$ \\
\end{longtable}}

{\setlength{\tabcolsep}{12pt}\renewcommand{\arraystretch}{1.6}\small
\begin{longtable}{c c c c c c}
\caption{Dirac quasinormal frequencies $\omega M$ for the potential $V_+$ (third-order WKB); $\ell$ labels $\kappa=\ell+1$.}\label{tab:qnmd}\\
\cellcolor{brown!30}$\alpha/M^2$ & \cellcolor{brown!30}$a/M$ & \cellcolor{brown!30}$\ell=1$ & \cellcolor{brown!30}$\ell=2$ & \cellcolor{brown!30}$\ell=3$ & \cellcolor{brown!30}$\ell=2,\,n=1$ \\ \hline\hline
\endfirsthead
\cellcolor{brown!30}$\alpha/M^2$ & \cellcolor{brown!30}$a/M$ & \cellcolor{brown!30}$\ell=1$ & \cellcolor{brown!30}$\ell=2$ & \cellcolor{brown!30}$\ell=3$ & \cellcolor{brown!30}$\ell=2,\,n=1$ \\ \hline\hline
\endhead
\hline\hline
\endfoot
0.00 & 0.00 & $0.3786-0.0965i$ & $0.5737-0.0963i$ & $0.7672-0.0963i$ & $0.5562-0.2930i$ \\
0.00 & 0.50 & $0.3788-0.0951i$ & $0.5738-0.0950i$ & $0.7672-0.0949i$ & $0.5570-0.2887i$ \\
0.00 & 1.00 & $0.3792-0.0908i$ & $0.5739-0.0907i$ & $0.7673-0.0907i$ & $0.5590-0.2754i$ \\
0.00 & 1.50 & $0.3793-0.0831i$ & $0.5739-0.0832i$ & $0.7673-0.0833i$ & $0.5601-0.2519i$ \\
0.50 & 0.00 & $0.3824-0.0942i$ & $0.5793-0.0942i$ & $0.7747-0.0942i$ & $0.5625-0.2862i$ \\
0.50 & 0.50 & $0.3826-0.0928i$ & $0.5793-0.0928i$ & $0.7747-0.0928i$ & $0.5632-0.2818i$ \\
0.50 & 1.00 & $0.3829-0.0885i$ & $0.5795-0.0885i$ & $0.7748-0.0885i$ & $0.5649-0.2684i$ \\
0.50 & 1.50 & $0.3829-0.0807i$ & $0.5794-0.0808i$ & $0.7748-0.0809i$ & $0.5655-0.2445i$ \\
1.00 & 0.00 & $0.3861-0.0912i$ & $0.5853-0.0914i$ & $0.7829-0.0914i$ & $0.5684-0.2773i$ \\
1.00 & 0.50 & $0.3863-0.0898i$ & $0.5854-0.0900i$ & $0.7830-0.0900i$ & $0.5691-0.2729i$ \\
1.00 & 1.00 & $0.3867-0.0854i$ & $0.5856-0.0856i$ & $0.7831-0.0857i$ & $0.5706-0.2592i$ \\
1.00 & 1.50 & $0.3868-0.0775i$ & $0.5855-0.0778i$ & $0.7830-0.0779i$ & $0.5706-0.2350i$ \\
1.50 & 0.00 & $0.3897-0.0872i$ & $0.5919-0.0875i$ & $0.7921-0.0876i$ & $0.5730-0.2655i$ \\
1.50 & 0.50 & $0.3900-0.0858i$ & $0.5920-0.0861i$ & $0.7922-0.0862i$ & $0.5736-0.2610i$ \\
1.50 & 1.00 & $0.3905-0.0814i$ & $0.5922-0.0817i$ & $0.7923-0.0818i$ & $0.5749-0.2473i$ \\
1.50 & 1.50 & $0.3906-0.0734i$ & $0.5922-0.0737i$ & $0.7923-0.0738i$ & $0.5743-0.2228i$ \\
\end{longtable}}

\begin{figure}[!htbp]
\centering
\includegraphics[width=0.92\textwidth]{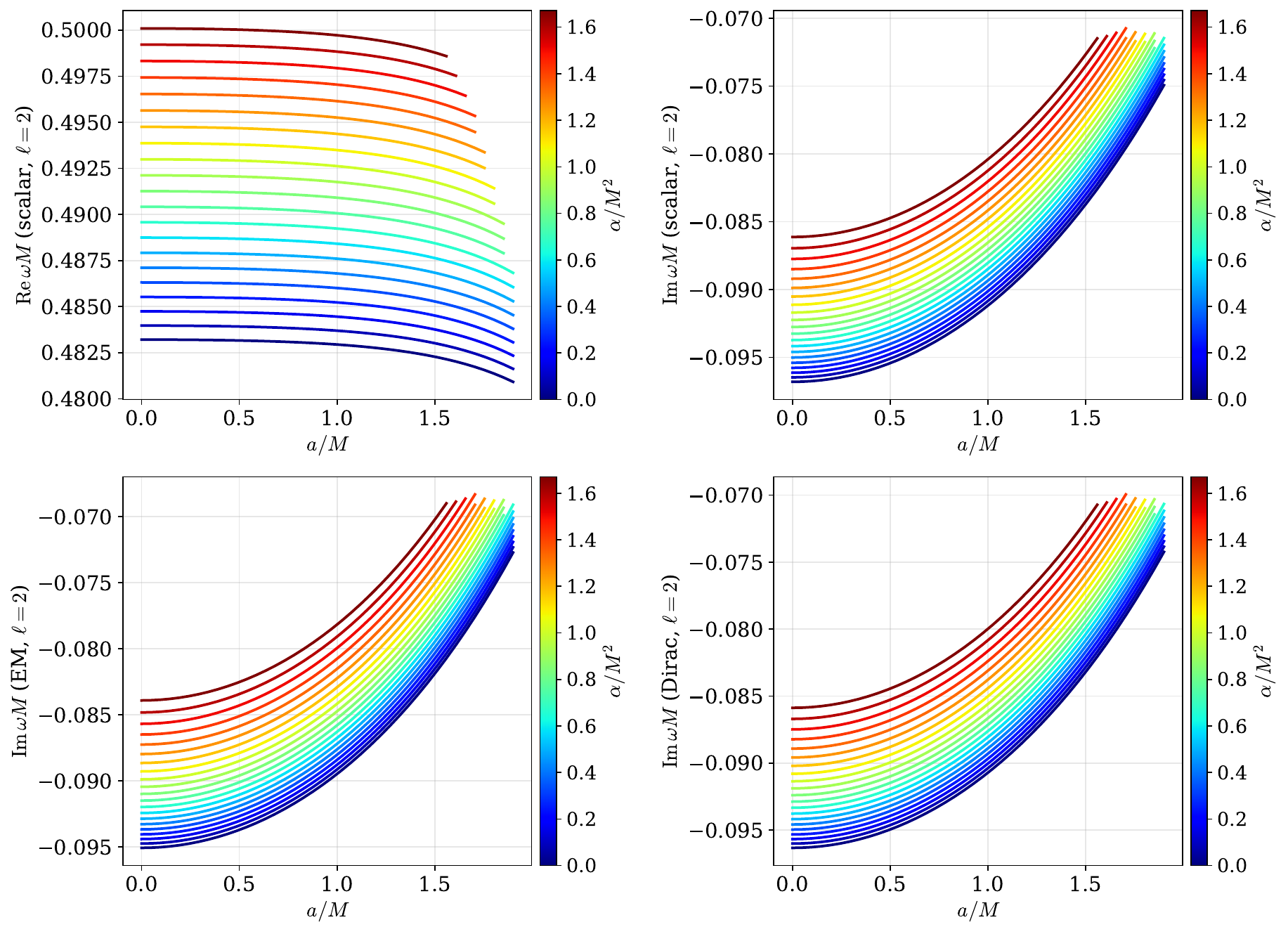}
\caption{Fundamental quasinormal frequencies from the third-order WKB formula as functions of the bounce parameter, for $\alpha$ from $0$ to $\alpha_{\rm ext}$ (color bar). Top left: real part of the scalar $\ell=2$ mode; top right: its imaginary part; bottom: imaginary parts of the electromagnetic and Dirac $\ell=2$ modes. Each ribbon ends where $a$ reaches $R_+(\alpha)$ and the horizon disappears.}
\label{fig:qnm}
\end{figure}

The pattern in Fig.~\ref{fig:qnm} is the one anticipated by Eqs.~(\ref{eq:bc}) and~(\ref{eq:lyap}). The real part of every mode is nearly flat in $a$, changing by less than $0.5\%$ between $a=0$ and $a=1.9\Ms$ for the scalar $\ell=2$ mode at $\alpha=0.5\Ms^2$, while it increases with $\alpha$, following the growth of $\Omega_c$. The imaginary part decreases in magnitude with $a$ at every $\alpha$, and it does so at the rate predicted by the eikonal factor: at the reference point the WKB damping is $0.0889/\Ms$ against $0.0946/\Ms$ at $a=0$, a ratio of $0.940$, and the factor $\sqrt{1-a^2/\Rps^2}$ of Table~\ref{tab:lam} is $0.9404$. The agreement holds to three digits already at $\ell=2$ and for all three spins, which is somewhat better than one might expect from a leading-order eikonal statement, and it is due to the fact that the $a$ dependence enters the potential only through the smooth factor $r/R$ in the derivative terms. The electromagnetic and Dirac modes are lower in frequency and slightly less damped than the scalar ones, in the usual order, and the overtones in the tables show that the suppression of the damping by the bounce grows with $n$, since higher overtones probe the potential further from its peak where $a/R$ is larger. The eikonal-correspondence caveat of~\cite{Konoplya:2017} is not an issue here: the potentials in Eqs.~(\ref{eq:Vs})-(\ref{eq:Vd}) all have the areal centrifugal term as their leading part and their peaks approach the photon sphere as $\ell\to\infty$.

To validate the WKB results and to obtain the $\ell=0$ modes we implemented the Chebyshev spectral collocation method of Batic, Dutykh and Sukaiti~\cite{Batic:2026,Batic:2026b}. The radial equation is written in the horizon-anchored variable $z=r/\rH$, the quasinormal asymptotics $\psi\sim(z-1)^{-i\omega/(2\kappa)}$ at the horizon and $\psi\sim e^{i\omega r_*}$ at infinity are factored out through $\psi=z^{2i\Omega(1+s_h)}(z-1)^{-2i\Omega s_h}e^{2ix_h\Omega(z-1)}\,\Phi(z)$ with $\Omega=\Ms\omega$, $x_h=\rH/2\Ms$ and $s_h=1/(2\Ms A'(\rH))$, the domain is compactified by $z=2/(1-y)$, and the regular function $\Phi$ is expanded in Chebyshev polynomials collocated at the Gauss roots. The result is a quadratic matrix pencil $(M_0+\Omega M_1+\Omega^2M_2)\vec a=0$, which we linearize and solve as a generalized eigenvalue problem in $30$-digit arithmetic; physical modes are those that agree across three resolutions $N=24,28,32$ within $10^{-4}$, and we report the average and the spread. On Schwarzschild the solver gives $\Omega=0.110457-0.104896i$ for the $\ell=0$ fundamental, against Leaver's $0.110455-0.104896i$~\cite{Leaver:1985}. Table~\ref{tab:spec} compares the spectral and WKB results on the bounce-qOSD geometry.

{\setlength{\tabcolsep}{10pt}\renewcommand{\arraystretch}{1.6}\small
\begin{longtable}{c c c c c c c}
\caption{Scalar quasinormal frequencies from the Chebyshev spectral collocation solver ($N=24,28,32$, 30-digit arithmetic) compared with the third-order WKB fundamental modes. The spread is the maximal difference between the three resolutions; the first overtone is listed where the solver resolved it. The first row is the Schwarzschild check against Leaver.}\label{tab:spec}\\
\cellcolor{brown!30}$\alpha/M^2$ & \cellcolor{brown!30}$a/M$ & \cellcolor{brown!30}$\ell$ & \cellcolor{brown!30}$n$ & \cellcolor{brown!30}spectral $\omega M$ & \cellcolor{brown!30}spread & \cellcolor{brown!30}WKB $\omega M$ \\ \hline\hline
\endfirsthead
\cellcolor{brown!30}$\alpha/M^2$ & \cellcolor{brown!30}$a/M$ & \cellcolor{brown!30}$\ell$ & \cellcolor{brown!30}$n$ & \cellcolor{brown!30}spectral $\omega M$ & \cellcolor{brown!30}spread & \cellcolor{brown!30}WKB $\omega M$ \\ \hline\hline
\endhead
\hline\hline
\endfoot
0.0 & 0.0 & 0 & 0 & $0.11046-0.10490i$ & $1e-05$ & $0.1046-0.1152i$ \\
0.5 & 0.0 & 0 & 0 & $0.11239-0.10208i$ & $9e-06$ & $0.1032-0.1104i$ \\
0.5 & 0.0 & 1 & 0 & $0.29604-0.09537i$ & $4e-09$ & $0.2938-0.0954i$ \\
0.5 & 0.0 & 1 & 1 & $0.26903-0.29777i$ & $6e-06$ & -- \\
0.5 & 0.0 & 2 & 0 & $0.48847-0.09461i$ & $1e-09$ & $0.4879-0.0946i$ \\
0.5 & 0.0 & 2 & 1 & $0.46982-0.28852i$ & $4e-07$ & -- \\
0.5 & 1.0 & 0 & 0 & $0.10994-0.09502i$ & $4e-05$ & $0.0968-0.1020i$ \\
0.5 & 1.0 & 1 & 0 & $0.29549-0.08965i$ & $6e-08$ & $0.2930-0.0895i$ \\
0.5 & 1.0 & 1 & 1 & $0.27138-0.27805i$ & $3e-05$ & -- \\
0.5 & 1.0 & 2 & 0 & $0.48821-0.08898i$ & $3e-09$ & $0.4876-0.0889i$ \\
0.5 & 1.0 & 2 & 1 & $0.47197-0.27060i$ & $6e-07$ & -- \\
0.5 & 1.5 & 2 & 0 & $0.48753-0.08145i$ & $2e-08$ & $0.4868-0.0814i$ \\
0.5 & 1.5 & 2 & 1 & $0.47200-0.24669i$ & $2e-06$ & -- \\
1.0 & 1.0 & 2 & 0 & $0.49333-0.08608i$ & $4e-09$ & $0.4926-0.0860i$ \\
1.0 & 1.0 & 2 & 1 & $0.47683-0.26115i$ & $7e-07$ & -- \\
\end{longtable}}

For $\ell\geq1$ the two methods agree to three or four digits, which confirms both the WKB numbers and the sign conventions of the spectral factorization on a metric whose lapse contains square roots. For $\ell=0$ the WKB values are off by about $10\%$ in the real part, as is standard for the lowest multipole, and the spectral values are the ones to be trusted. The spectral scan also returned no eigenvalue with a positive imaginary part that survived the three-resolution filter, for any of the configurations examined, so the black-hole members of the family are linearly stable against scalar perturbations in the range explored. The first overtone found by the solver, $\Omega=0.2690-0.2978i$ for $\ell=1$ at $a=0$, agrees with the WKB overtone to within the expected accuracy of the latter.

The third method is direct time-domain integration~\cite{Gundlach:1994}, which we use both as an independent check of the fundamental mode and to display the wormhole echoes. We integrate $\partial_t^2\psi-\partial_{r_*}^2\psi+V\psi=0$ on the characteristic grid $u=t-r_*$, $v=t+r_*$ with the standard second-order stencil, starting from a Gaussian pulse of width $3\Ms$ on the initial null surface, and read off the signal at $r_*=15\Ms$. For the black holes the tortoise coordinate is obtained by quadrature from the horizon; for the wormholes from the throat. A fit of the late-time envelope and zero crossings at the reference point gives $\omega\Ms=0.488-0.089i$, in agreement with the WKB value $0.4876-0.0889i$ and with the spectral value.

\begin{figure}[!htbp]
\centering
\includegraphics[width=0.95\textwidth]{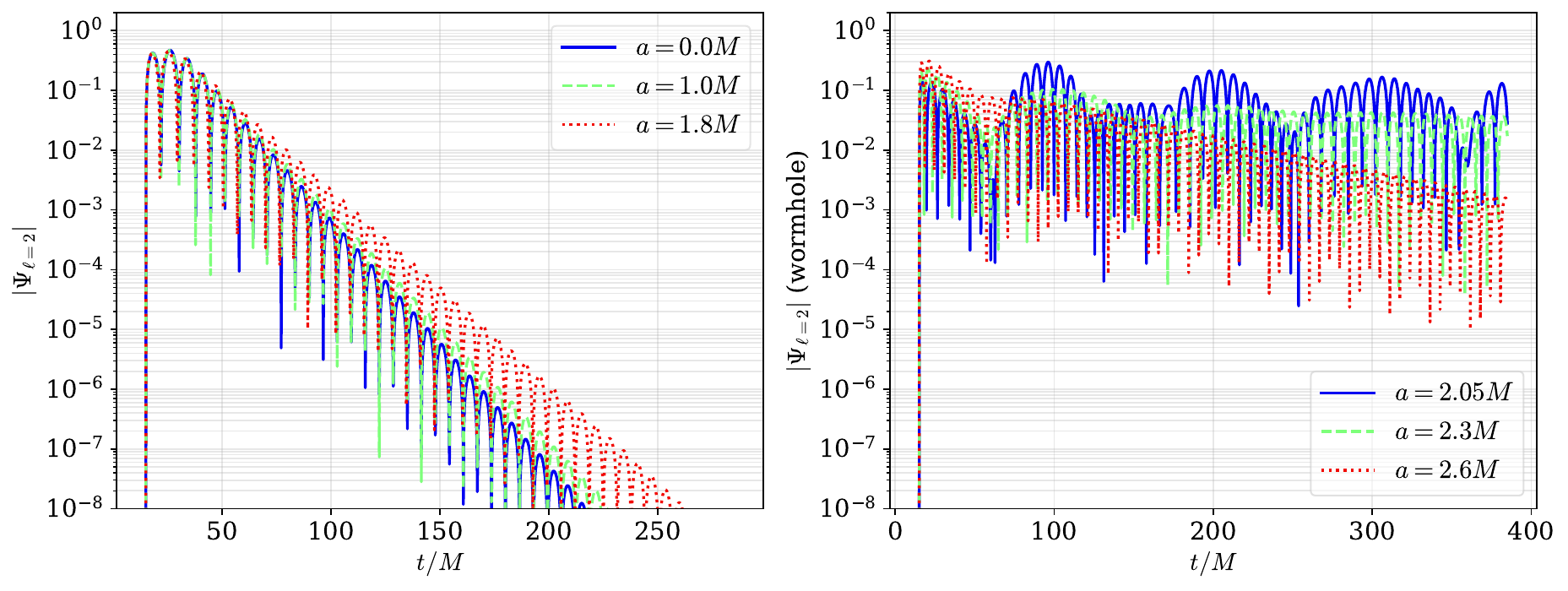}
\caption{Time-domain profiles of the scalar $\ell=2$ field at $\alpha=0.5M^2$. Left: black-hole configurations with $a=0$, $M$ and $1.8M$; the ringdown becomes longer as $a$ grows, and the late-time power-law tail is the same. Right: wormhole configurations with $a=2.05M$, $2.3M$ and $2.6M$, just above $R_+=1.93M$; the prompt ringdown of the outer barrier is followed by a train of echoes from the well between the two peaks of Fig.~\ref{fig:pot}, whose spacing shrinks as $a$ increases and the well narrows.}
\label{fig:td}
\end{figure}

Figure~\ref{fig:td} shows the profiles. In the black-hole regime the three signals share the same oscillation period, as the real part of the mode is insensitive to $a$, and differ in their decay, the $a=1.8\Ms$ signal being visibly longer-lived, exactly as the ribbons of Fig.~\ref{fig:qnm} say. All three settle into the same power-law tail, which is governed by the asymptotic structure of the potential and not by the bounce. This is expected from Price's analysis~\cite{Gundlach:1994}: the late-time decay $\psi\sim t^{-(2\ell+3)}$ is fixed by the leading $1/r$ falloff of the lapse, which is the same $2M/r$ for every member of the family, and the $a^2/r^3$ and $\alpha/r^4$ corrections only modify subleading terms. The bounce therefore leaves an imprint on the intermediate ringdown stage, through the damping rate, and none on either the prompt response, which is set by the initial data, or the tail. In terms of the three-stage picture of a perturbed black hole this places the bounce parameter squarely in the quasinormal window, which is where it should be looked for. In the wormhole regime the picture changes qualitatively. The initial burst rings down at the frequency of the outer barrier, but after a delay set by twice the tortoise distance between the two peaks the trapped part of the wave leaks out as a first echo, and further echoes follow at regular intervals with slowly decaying amplitude. The echo delay is longest for $a=2.05\Ms$, where the throat is barely outside the would-be horizon, $A(a)$ is small and the well is deep and wide in $r_*$, and it decreases for $a=2.3\Ms$ and $2.6\Ms$; at the same time the echoes become less distinct because the barriers shrink and the trapping weakens. This is the behavior found for the Simpson-Visser wormhole~\cite{Churilova:2020,Yang:2023b,Ou:2022} and for the brane-world transition of~\cite{Bronnikov:2022}, and it is the wave-optics counterpart of the ray-and-wave analysis we carried out for lower-dimensional charged wormholes~\cite{Ahmed:2026c}; the holonomy correction does not alter it beyond shifting the location of the transition to $R_+(\alpha)<2\Ms$. The delay can be estimated from the geometry alone: successive echoes are separated by twice the tortoise distance between the two barriers, $\Delta t_{\rm echo}\simeq2\int_{-\rps}^{\rps}\dd r/A$, which gives $72.6M$, $36.2M$ and $20.4M$ for $a=2.05M$, $2.3M$ and $2.6M$ respectively, in agreement with the spacing of the bursts in the right panel of Fig.~\ref{fig:td}. The delay grows without bound as $a\to R_+^+$, where $A(a)\to0$ and the throat becomes a null surface, which is the black-hole-to-wormhole transition studied in~\cite{Churilova:2020,Bronnikov:2022}; in that limit the echoes are separated by an arbitrarily long quiet interval, and the prompt ringdown alone is indistinguishable from that of the one-way wormhole. The wormhole modes themselves are long-lived quasibound states of the double-barrier potential and can be obtained with the spectral solver on the full axis, a task we leave for future work.

The spectral solver also resolves the first two overtones of the $\ell=2$ scalar mode, listed in Table~\ref{tab:spec}, and they display the same decoupling as the fundamental. At $\alpha=0.5M^2$ the ratios ${\rm Im}\,\omega_n/{\rm Im}\,\omega_0$ are $3.05$ and $5.23$ at $a=0$ and $3.04$ and $5.19$ at $a=M$, so the overtone ladder keeps its nearly equal spacing of $\Delta({\rm Im}\,\omega)\simeq2|{\rm Im}\,\omega_0|$ expected from the eikonal formula $\omega_n\simeq\Omega_c\ell-i(n+\tfrac12)\lambda_L$, while the whole ladder is compressed by the bounce factor; the real parts decrease along the ladder by $4\%$ and $10\%$ at $a=0$ and by $3\%$ and $9\%$ at $a=M$. The spread across resolutions grows from $10^{-9}$ for the fundamental to $10^{-5}$ for the second overtone, which is the usual loss of accuracy of collocation methods for strongly damped modes and could be reduced by increasing $N$ and the working precision~\cite{Batic:2026}.

\section{Greybody bounds, Hawking emission and sparsity}\label{isec7}

We now quantify what fraction of the Hawking flux escapes the potential barrier and how the bounce reshapes the emission spectrum. Rather than integrating the wave equation for each frequency we use the exact lower bound on the transmission probability derived by Visser~\cite{Visser:1999} and applied to black holes by Boonserm and Visser~\cite{Boonserm:2008},
\begin{equation}
T_\ell(\omega)\;\geq\;T_b^{(\ell)}(\omega)={\rm sech}^2\!\left[\frac{1}{2\omega}\int_{-\infty}^{\infty}V_\ell\,\dd r_*\right]
={\rm sech}^2\!\left[\frac{1}{2\omega}\int\frac{V_\ell(r)}{A(r)}\,\dd r\right],
\label{eq:Tb}
\end{equation}
where the $r$ integral runs from $\rH$ to infinity for a black hole and over the whole axis for a wormhole. The bound depends on the potential only through the barrier integral $I_\ell=\int V_\ell\dd r_*$, which we evaluate numerically; it is exact in the sense of being a true inequality, and it captures the low-frequency suppression and the high-frequency saturation correctly~\cite{Rincon:2020,AlBadawi:2023,AlBadawi:2024,Lutfuoglu:2026c}. For Schwarzschild the scalar $\ell=0$ integral is $I_0=1/(4\Ms)$, and our code returns $0.2500/\Ms$ at $\alpha=a=0$. Table~\ref{tab:gb} lists $I_\ell$ and $T_b^{(\ell)}$ at $\omega\Ms=0.3$ for the reference $\alpha$ and a range of $a$, including one wormhole configuration.

{\setlength{\tabcolsep}{12pt}\renewcommand{\arraystretch}{1.6}\small
\begin{longtable}{c c c c c c c}
\caption{Barrier integrals $I_\ell=\int V_\ell\,dr_*$ of the scalar potential and the corresponding lower bounds $T_b$ on the transmission probability at $\omega M=0.3$, for $\alpha=0.5M^2$. The last row ($a=2.3M$) is a traversable wormhole, for which the integral runs over both sides of the throat.}\label{tab:gb}\\
\cellcolor{brown!30}$a/M$ & \cellcolor{brown!30}$I_0 M$ & \cellcolor{brown!30}$T_b^{(0)}$ & \cellcolor{brown!30}$I_1 M$ & \cellcolor{brown!30}$T_b^{(1)}$ & \cellcolor{brown!30}$I_2 M$ & \cellcolor{brown!30}$T_b^{(2)}$ \\ \hline\hline
\endfirsthead
\cellcolor{brown!30}$a/M$ & \cellcolor{brown!30}$I_0 M$ & \cellcolor{brown!30}$T_b^{(0)}$ & \cellcolor{brown!30}$I_1 M$ & \cellcolor{brown!30}$T_b^{(1)}$ & \cellcolor{brown!30}$I_2 M$ & \cellcolor{brown!30}$T_b^{(2)}$ \\ \hline\hline
\endhead
\hline\hline
\endfoot
0.00 & 0.2534 & 0.84088 & 1.2861 & 0.05351 & 3.3514 & 0.0000563 \\
0.50 & 0.2520 & 0.84243 & 1.2966 & 0.05171 & 3.3858 & 0.0000502 \\
1.00 & 0.2477 & 0.84719 & 1.3334 & 0.04588 & 3.5048 & 0.0000338 \\
1.50 & 0.2401 & 0.85555 & 1.4233 & 0.03421 & 3.7896 & 0.0000131 \\
1.90 & 0.2307 & 0.86562 & 1.6934 & 0.01404 & 4.6190 & 0.0000008 \\
2.30 & 0.4324 & 0.61888 & 3.1576 & 0.00011 & 8.6079 & 0.0000000 \\
\end{longtable}}

\begin{figure}[!htbp]
\centering
\includegraphics[width=\textwidth]{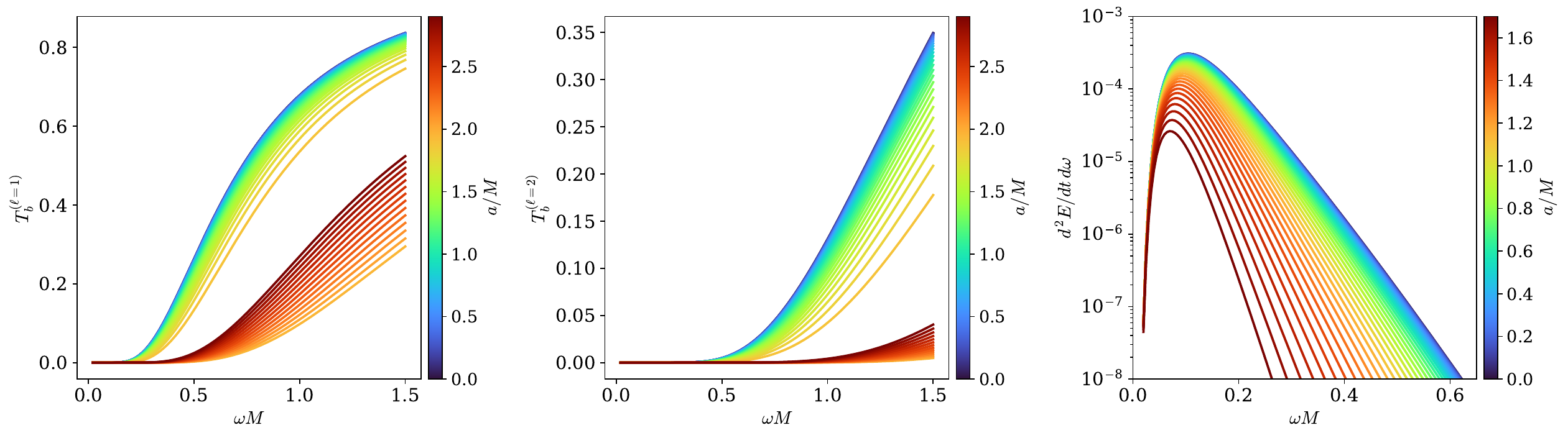}
\caption{Lower bounds on the scalar transmission probability for $\ell=1$ (left) and $\ell=2$ (middle) at $\alpha=0.5M^2$, as $a$ runs from $0$ to $2.9M$; the black-hole curves (blue to yellow) shift slightly to higher frequency with $a$, while the wormhole curves (orange to red) are strongly suppressed because the barrier integral covers both sides of the throat. Right: Hawking energy emission rate of the scalar field summed over $\ell\leq3$, for black-hole configurations with $a$ from $0$ to $1.7M$; the peak moves to lower frequency and lower height as the temperature drops with $a$.}
\label{fig:gb}
\end{figure}

Two effects are visible in Fig.~\ref{fig:gb} and Table~\ref{tab:gb}. In the black-hole regime the barrier integral $I_\ell$ grows slowly with $a$, by $2\%$ for $\ell=1$ between $a=0$ and $a=1.5\Ms$, because the bounce term in Eq.~(\ref{eq:Vs}) is positive and because the factor $1/A$ in the measure gives more weight to the near-horizon region that the bounce widens in $r_*$; the transmission is therefore somewhat lower at fixed frequency. For $\ell=0$ the integral decreases instead, since the $rA'/R^2$ term is reduced when $r<R$, and the $\ell=0$ bound improves slightly with $a$. Crossing into the wormhole regime changes $I_\ell$ by a large factor, $2.4$ for $\ell=2$ at $a=2.3\Ms$, because the wave has to tunnel through two barriers, and the transmission at $\omega\Ms=0.3$ drops by seven orders of magnitude. The Hawking spectrum of the black holes, computed from the Page formula $\dd^2E/\dd t\,\dd\omega=(2\pi)^{-1}\sum_\ell(2\ell+1)T_b^{(\ell)}\,\omega/(e^{\omega/T_H}-1)$ with the bounds inserted for the transmission~\cite{Hawking:1975,Page:1976}, is dominated by the temperature. Since $T_H$ falls from $0.0368/\Ms$ at $a=0$ to $0.0174/\Ms$ at $a=1.7\Ms$ for $\alpha=0.5\Ms^2$, the peak of the emission moves from $\omega\Ms\simeq0.105$ at $a=0$ through $0.097$ at the reference point to about $0.05$ at $a=1.7\Ms$, and its height drops by more than an order of magnitude over the same range. A bounce-regularized black hole therefore evaporates more slowly than its qOSD seed, and the evaporation stops altogether in the limit $a\to R_+$ where $T_H\to0$; this is the endpoint discussed for evaporating black bounces in~\cite{Lutfuoglu:2026a}, and the holonomy correction does not change the qualitative conclusion.

The sparsity of the Hawking cascade, introduced by Gray, Schuster, Van-Brunt and Visser~\cite{Gray:2016}, measures whether the emitted quanta are separated in time by more than their own wavelength. With the thermal wavelength $\lambda_{\rm th}=2\pi/T_H$ and the effective emitting area $A_{\rm eff}=4\pi b_c^2$ appropriate to the geometric-optics cross section one has
\begin{equation}
\eta=\frac{\lambda_{\rm th}^2}{A_{\rm eff}}=\frac{\pi}{T_H^2\,b_c^2},\qquad \eta_{\rm Schw}=\frac{64\pi^3}{27}\simeq73.5 ,
\label{eq:eta}
\end{equation}
and our numerical value at $\alpha=a=0$ reproduces $73.50$. The ringdown quality factor $Q={\rm Re}\,\omega/(2|{\rm Im}\,\omega|)$ complements the sparsity as a measure of how many oscillations a mode completes before decaying, and it is the quantity that determines whether a mode can be resolved in a gravitational-wave observation~\cite{Berti:2009,Cardoso:2019}.

\begin{figure}[!htbp]
\centering
\includegraphics[width=0.95\textwidth]{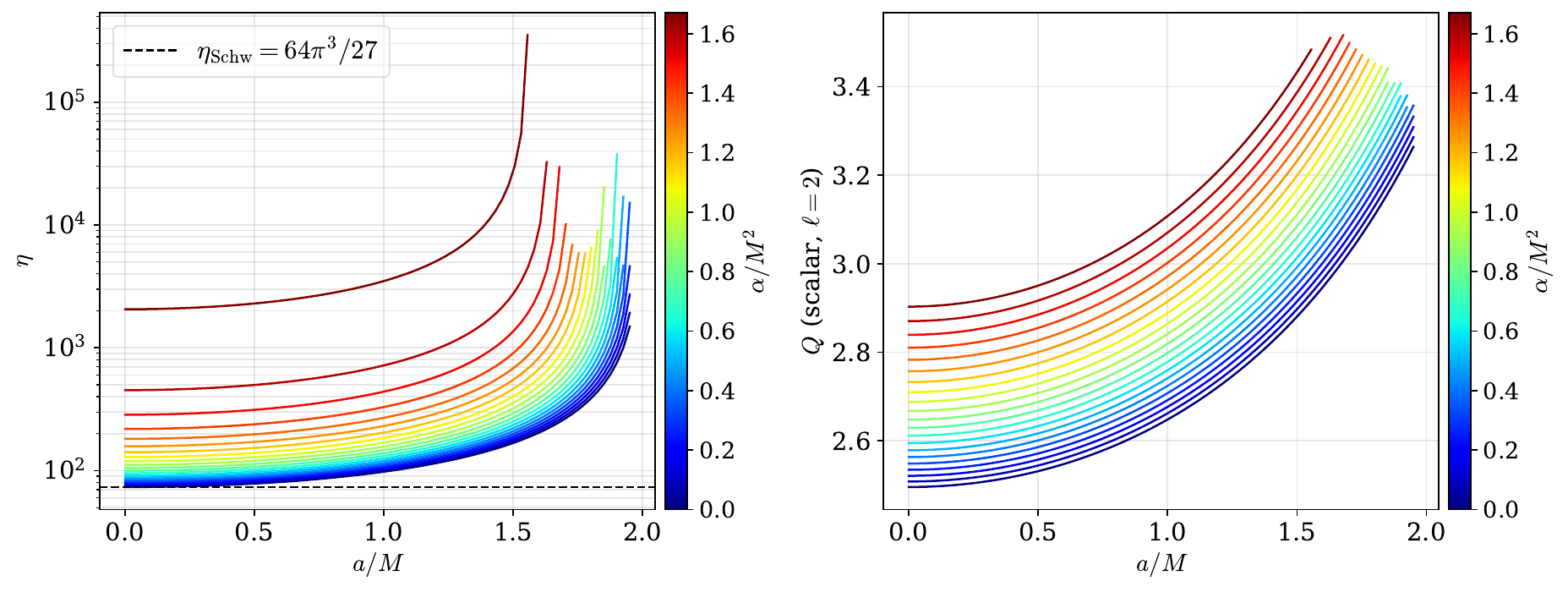}
\caption{Left: sparsity of the Hawking cascade, Eq.~(\ref{eq:eta}), as a function of $a$ for $\alpha$ from $0$ to $\alpha_{\rm ext}$, on a logarithmic scale, with the Schwarzschild value dashed. Right: quality factor of the fundamental scalar $\ell=2$ mode on the same grid. Both quantities increase with $a$ and with $\alpha$ and diverge as $a\to R_+(\alpha)$.}
\label{fig:sp}
\end{figure}

{\setlength{\tabcolsep}{12pt}\renewcommand{\arraystretch}{1.6}\small
\begin{longtable}{c c c c c c}
\caption{Hawking temperature, capture impact parameter, sparsity $\eta$ of the Hawking cascade and quality factor $Q$ of the fundamental $\ell=2$ scalar mode.}\label{tab:sp}\\
\cellcolor{brown!30}$\alpha/M^2$ & \cellcolor{brown!30}$a/M$ & \cellcolor{brown!30}$T_H M$ & \cellcolor{brown!30}$b_c/M$ & \cellcolor{brown!30}$\eta$ & \cellcolor{brown!30}$Q$ \\ \hline\hline
\endfirsthead
\cellcolor{brown!30}$\alpha/M^2$ & \cellcolor{brown!30}$a/M$ & \cellcolor{brown!30}$T_H M$ & \cellcolor{brown!30}$b_c/M$ & \cellcolor{brown!30}$\eta$ & \cellcolor{brown!30}$Q$ \\ \hline\hline
\endhead
\hline\hline
\endfoot
0.00 & 0.00 & 0.03979 & 5.1962 & 73.50 & 2.496 \\
0.00 & 0.50 & 0.03853 & 5.1962 & 78.40 & 2.532 \\
0.00 & 1.00 & 0.03446 & 5.1962 & 98.00 & 2.648 \\
0.00 & 1.50 & 0.02632 & 5.1962 & 167.99 & 2.881 \\
0.50 & 0.00 & 0.03677 & 5.1459 & 87.75 & 2.579 \\
0.50 & 0.50 & 0.03551 & 5.1459 & 94.06 & 2.617 \\
0.50 & 1.00 & 0.03145 & 5.1459 & 119.93 & 2.741 \\
0.50 & 1.50 & 0.02315 & 5.1459 & 221.45 & 2.992 \\
1.00 & 0.00 & 0.03192 & 5.0904 & 118.96 & 2.688 \\
1.00 & 0.50 & 0.03072 & 5.0904 & 128.45 & 2.729 \\
1.00 & 1.00 & 0.02679 & 5.0904 & 168.88 & 2.864 \\
1.00 & 1.50 & 0.01847 & 5.0904 & 355.21 & 3.138 \\
\end{longtable}}

Figure~\ref{fig:sp} and Table~\ref{tab:sp} show that both quantities increase monotonically with each parameter. The sparsity is $\eta=120$ at the reference point and $\eta=355$ at $(\alpha,a)=(\Ms^2,1.5\Ms)$, so the cascade is even sparser than in the Schwarzschild case, which was already extremely sparse; the emission is a sequence of well-separated quanta rather than a thermal fluid, and the bounce reinforces this by lowering $T_H$ while leaving $b_c$ untouched. The quality factor grows from $2.50$ for Schwarzschild to $2.74$ at the reference point and to $3.14$ at $(\Ms^2,1.5\Ms)$, because the real part stays put while the damping is reduced. The ratio $Q/Q_{\rm Schw}$ is, to within a percent, the inverse of the suppression factor $\sqrt{1-a^2/\Rps^2}$ times the qOSD ratio, which is another way of stating the decoupling of Sec.~\ref{isec4}: the bounce makes the black hole a sharper resonator without changing its pitch.
\section{Observational outlook: which observable sees which parameter}\label{isec7b}

We collect here the observational content of the previous sections, since the decoupling between the two parameters is the feature of the family most likely to be useful in practice. The rule is simple. Any quantity that is defined by the areal lapse $A(R)$ and by the position of a circular orbit, without reference to derivatives along the radial proper direction, depends on $\alpha$ alone; any quantity that involves the curvature of an effective potential along the proper radial direction acquires one factor of $\sqrt{1-a^2/R^2}$ at the orbit in question. Table~\ref{tab:obs} lists the observables of this paper in that light, together with the value of each quantity at the reference point $(\alpha,a)=(0.5M^2,M)$ divided by its Schwarzschild value, so that a ratio below unity means suppression and a ratio above unity enhancement.

{\setlength{\tabcolsep}{12pt}\renewcommand{\arraystretch}{1.6}\small
\begin{longtable}{l c c c}
\caption{Parameter dependence of the observables studied in this paper and the ratio of each quantity at the reference point $(\alpha,a)=(0.5M^2,M)$ to its Schwarzschild value. The middle columns state whether the quantity depends on the holonomy parameter and on the bounce parameter.}\label{tab:obs}\\
\cellcolor{brown!30}observable & \cellcolor{brown!30}$\alpha$ & \cellcolor{brown!30}$a$ & \cellcolor{brown!30}ratio to Schwarzschild at reference point \\ \hline\hline
\endfirsthead
\cellcolor{brown!30}observable & \cellcolor{brown!30}$\alpha$ & \cellcolor{brown!30}$a$ & \cellcolor{brown!30}ratio to Schwarzschild at reference point \\ \hline\hline
\endhead
\hline\hline
\endfoot
shadow radius $b_c$ & yes & no & $0.990$ \\
photon-sphere frequency $\Omega_c$, ${\rm Re}\,\omega$ (eikonal) & yes & no & $1.010$ \\
Lyapunov exponent $\lambda_L$, ${\rm Im}\,\omega$ (eikonal) & yes & yes & $0.92$ \\
${\rm Re}\,\omega$ (scalar $\ell=2$, WKB) & yes & weak & $1.009$ \\
${\rm Im}\,\omega$ (scalar $\ell=2$, WKB) & yes & yes & $0.92$ \\
quality factor $Q$ & yes & yes & $1.10$ \\
ISCO areal radius, $E_{\rm ISCO}$, $L_{\rm ISCO}$ & yes & no & $0.987$ (radius) \\
Keplerian frequency $\nu_\phi$, vertical $\nu_\theta$ & yes & no & $1.013$ at $R=8M$ \\
radial epicyclic frequency $\nu_r$ & yes & yes & $0.987$ at $R=8M$ \\
$3{:}2$ resonance radius $R_{3:2}$ & yes & yes & $1.003$ \\
Hawking temperature $T_H$ & yes & yes & $0.79$ \\
first-law entropy at fixed $R_h=2M$ & yes & yes & $1.42$ \\
sparsity $\eta$ & yes & yes & $1.63$ \\
greybody barrier integral $I_1$ & yes & yes & $0.68$ \\
echo delay (wormhole regime) & yes & yes & absent (needs $a>R_+$) \\
proper time horizon to throat & yes & yes & finite ($1.6M$, no counterpart) \\
\end{longtable}}

Three conclusions follow from the table. The first is that the shadow is a poor probe of this family. The ratio $b_c/b_c^{\rm Schw}$ stays above $0.96$ over the entire physical range of $\alpha$, well inside the present Sgr~A$^*$ bounds, and $a$ does not enter at all; the analyses of the qOSD shadow in~\cite{Yang:2023,Luo:2024,Yang:2026} therefore transfer unchanged to the regularized geometry, and neither the EHT nor its planned successors will separate a bounce-regularized qOSD black hole from Schwarzschild through the shadow alone. The second is that the ringdown does better, and does so through the damping rather than the frequency. At the reference point the fundamental $\ell=2$ mode of every spin has its real part raised by a factor close to $1.01$, an amount set by $\alpha$ and comparable to the shadow shift, while its damping rate is multiplied by $0.92$, of which the factor $0.94$ comes from the bounce; the quality factor increases accordingly. Since the damping time of the dominant mode of a binary merger remnant is measured at the ten-percent level already~\cite{Abbott:2016,Cardoso:2019}, and the frequency far better, a joint fit of frequency and damping isolates the bounce factor $\sqrt{1-a^2/\Rps^2}$ directly, provided the test-field results carry over to the gravitational sector, which is the open task named in the conclusions. The third is that the epicyclic frequencies carry a clean but small signature: the Keplerian and vertical frequencies see $\alpha$, the radial one sees both, and a simultaneous measurement of the upper and lower peaks in a source of known mass returns $a$ through the ratio $\nu_r/\nu_\phi$ with no leading-order degeneracy with $\alpha$; for $a\sim M$ the factor on $\nu_r$ near the ISCO is about $0.99$, a shift at the edge of what the current twin-peak data allow~\cite{Motta:2014,Stella:1999}.

The thermodynamic entries in the table are not observables in the astrophysical sense, but they are where the two parameters act most strongly, and they matter for primordial or microscopic black holes. A bounce-regularized qOSD black hole of given mass is colder, more entropic and sparser in its Hawking emission than its Schwarzschild counterpart, and it stops radiating altogether when the throat reaches the horizon, leaving a zero-temperature one-way wormhole as the endpoint of evaporation; this connects the present family to the evaporation scenarios of~\cite{Lutfuoglu:2026a,Calza:2025} and to the question of black-hole remnants as dark-matter candidates. The wormhole sector, finally, is observable only through echoes, whose delay is set by the tortoise width of the well and grows without bound at the transition; a ringdown followed by a quiet interval and a late burst would be the signature, and its absence in a given event bounds $a$ from above only if $a$ is close enough to $R_+$ for the delay to fall within the observation window.
\section{Conclusions}\label{isec8}

We have constructed and analyzed a two-parameter family of static, spherically symmetric geometries that combines the holonomy-corrected exterior of the quantum Oppenheimer-Snyder-Datt collapse with the Simpson-Visser bounce. The seed lapse $1-2M/R+\alpha M^2/R^4$ is kept as a function of the areal radius, and the areal radius is replaced by $\sqrt{r^2+a^2}$. The family reduces to the qOSD black hole, to the Simpson-Visser black bounce, and to Schwarzschild in the appropriate limits, and every result of the paper was checked against these limits, symbolically where the expressions are closed and numerically otherwise. The geometry has finite Ricci and Kretschmann scalars for every $a>0$, with the curious property that the Ricci scalar at the throat, $2(3M-a)/a^3$, does not know about $\alpha$ at all, while the Kretschmann scalar retains the correction through terms that scale as $a^{-12}$. The effective source violates the null energy condition at the throat by the universal amount $T_{\mu\nu}k^\mu k^\nu=-1/(4\pi a^2)$, independent of $\alpha$ and $M$, so the holonomy correction neither causes nor cures the exotic matter that supports the bounce.

The parameter plane separates into a two-horizon regular black hole, a single-horizon black hole with a hidden bounce, and a traversable wormhole, with boundaries at $a=R_-(\alpha)$ and $a=R_+(\alpha)$ and with $\alpha_{\rm ext}=27M^2/16$ as the upper edge of the black-hole sector. We argued that the hidden-bounce regime is the physically interesting one: the inner horizon of the qOSD solution, and with it the mass-inflation channel that afflicts regular black holes, is removed from the manifold rather than regularized, the spacetime keeps a single event horizon, and the curvature stays bounded. The bounce achieves in one step what requires either fine-tuning to a degenerate inner horizon or a modified core in other constructions.

Horizon thermodynamics follows once the surface gravity is computed with the correct factor $\sqrt{1-a^2/R_h^2}$. The temperature vanishes both at the extremal areal radius $R_0=\sqrt{4\alpha/3}$ and at the throat-horizon coincidence $R_h=a$, the first-law entropy exceeds the area law by an amount that grows with both parameters and is elementary at $\alpha=0$, and the heat capacity has a Davies point at every $a$. The thermodynamic topological charge is $W=0$ throughout the black-hole sector. For $\alpha>0$ this is the charge the qOSD seed already had; for $\alpha=0$ it means that any bounce, however small, moves Schwarzschild from the $W=-1$ class to the $W=0$ class, and we identified the zero of the temperature at $R_h=a$ as the mechanism, with the minimum of the defect curve at $R_h=\sqrt2\,a$ and $\tau_{\min}=8\pi a$ in the Simpson-Visser limit.

The single most useful structural fact of the family is that circular geodesics are governed by the areal lapse. The photon-sphere radius, the critical impact parameter, the shadow, the ISCO and the Keplerian frequency depend on $\alpha$ only, whereas the Lyapunov exponent and the radial epicyclic frequency carry the exact factor $\sqrt{1-a^2/R^2}$. We verified this scaling to machine precision and then saw it reappear, to three digits, in the ratio of WKB damping rates at $\ell=2$ for scalar, electromagnetic and Dirac fields. The observational consequence is a clean division of labor: the shadow tests $\alpha$ and is in fact unable to bound it, since the impact parameter varies by under four percent over the entire physical range and stays inside the Sgr~A$^*$ band; the ringdown damping time, the quality factor, and the periastron-precession frequency of the relativistic precession model test $a$, with no leading-order degeneracy with $\alpha$. The twin-peak $3{:}2$ resonance radius moves outward with $a$ and inward with $\alpha$, so the two parameters could in principle be separated with a simultaneous measurement of the two peaks in a source of known mass.

For the spectra themselves we used three methods that agree with one another. Third-order WKB gives the fundamental modes and first overtones for $\ell\geq1$; Chebyshev spectral collocation, validated against Leaver on Schwarzschild, gives the $\ell=0$ modes and confirms the WKB numbers to three or four digits at higher $\ell$, finding no unstable mode; and characteristic time-domain integration reproduces the reference-point fundamental to the third digit and displays the echoes of the wormhole regime, where the potential develops two barriers and the trapped wave leaks out in a train of pulses whose spacing shrinks as $a$ increases. The Visser bound on the transmission probability shows that the black-hole barrier integrals move by a few percent with $a$ while the wormhole barrier integrals grow by a factor of two or more, that the Hawking emission peak moves to lower frequency and lower power as the bounce lowers the temperature, and that the cascade becomes even sparser than in the Schwarzschild case, with $\eta$ rising from $73.5$ to $120$ at the reference point.

In the near future we intend to extend the analysis in four directions. The first is the spectrum of the wormhole sector on the full axis, where the double-barrier potential supports long-lived quasibound modes that the spectral solver can capture once the factorization is adapted to two asymptotic regions; the echo delays measured here provide the target. The second is the axial gravitational sector with an effective source, following the treatment used for other non-vacuum regular geometries, which would allow a direct comparison with the ringdown of binary mergers rather than with test fields. The third is a rotating generalization through the Azreg-A\"inou or Newman-Janis route applied to the areal form of the metric, so that the decoupling between $\alpha$ and $a$ can be re-examined in the presence of frame dragging, where the nodal precession no longer vanishes and the relativistic precession model regains its third frequency. The fourth is a Bayesian confrontation of the epicyclic frequencies with the twin-peak data of the microquasars, which is the observational channel in which the bounce parameter is least degenerate and which the present computation has prepared.

\section*{Acknowledgments}

F.A. would like to thank the Inter-University Centre for Astronomy and Astrophysics (IUCAA), Pune, India for granting a visiting associateship. \.{I}.S. would like to acknowledge networking support of COST Actions CA18108, CA22113, CA21106, CA23130, CA23115, and CA21136. He also thanks EMU, T\"UB\.{I}TAK, ANKOS, and SCOAP3 for their support.

\section*{Data availability statement}

All data generated in this work are contained in the figures and tables of the article. The Python scripts used for the symbolic verification of the limits, the WKB and spectral quasinormal-mode computations, the time-domain integration and the figures are available from the corresponding author upon reasonable request.

\appendix
\section{Action and field equations supporting the geometry}\label{isecA}

The metric~(\ref{eq:metric}) was introduced as a geometric regularization, and the effective source of Sec.~\ref{isec2} was read off from the Einstein tensor. Here we show that the same source follows from a Lagrangian field theory, so that Eq.~(\ref{eq:metric}) is an exact solution of a definite set of field equations rather than a metric with an assigned stress tensor. We follow the construction that Bronnikov and Walia used for the Simpson-Visser spacetime~\cite{Bronnikov:2022}: a self-interacting scalar field with reversed kinetic sign together with a nonlinear electrodynamics carrying a magnetic charge. We use the signature $(-,+,+,+)$, the curvature convention $R^\rho{}_{\sigma\mu\nu}=\partial_\mu\Gamma^\rho_{\nu\sigma}-\partial_\nu\Gamma^\rho_{\mu\sigma}+\Gamma^\rho_{\mu\lambda}\Gamma^\lambda_{\nu\sigma}-\Gamma^\rho_{\nu\lambda}\Gamma^\lambda_{\mu\sigma}$ with $R_{\mu\nu}=R^\rho{}_{\mu\rho\nu}$, and the Einstein equations in the form $G_{\mu\nu}=8\pi T_{\mu\nu}$ with $G=c=1$. The action is
\begin{equation}
I=\int\dd^4x\sqrt{-g}\left[\frac{\mathcal{R}}{16\pi}-\frac{\epsilon}{2}\,g^{\mu\nu}\partial_\mu\phi\,\partial_\nu\phi-V(\phi)-\mathcal{L}(\mathcal{F})\right]
+\frac{1}{8\pi}\oint_{\partial\mathcal{M}}\dd^3y\sqrt{|h|}\,K,
\qquad \mathcal{F}=\frac14F_{\mu\nu}F^{\mu\nu},
\label{eq:action}
\end{equation}
where $\epsilon=\pm1$ selects a canonical or a phantom scalar, $F_{\mu\nu}=\partial_\mu A_\nu-\partial_\nu A_\mu$, and the Gibbons-Hawking-York term makes the variational problem well posed with Dirichlet data on the boundary; it does not affect the bulk equations but is needed if the on-shell action is used for thermodynamics.

Varying with respect to the inverse metric, using $\delta\sqrt{-g}=-\tfrac12\sqrt{-g}\,g_{\mu\nu}\delta g^{\mu\nu}$ and the Palatini identity $\delta\mathcal{R}=R_{\mu\nu}\delta g^{\mu\nu}+\nabla_\sigma v^\sigma$, whose divergence is cancelled by the boundary term, gives $G_{\mu\nu}=8\pi(T^{(\phi)}_{\mu\nu}+T^{(F)}_{\mu\nu})$ with the Hilbert tensors $T_{\mu\nu}=-2\,\partial\mathcal{L}_{\rm m}/\partial g^{\mu\nu}+g_{\mu\nu}\mathcal{L}_{\rm m}$,
\begin{equation}
T^{(\phi)}_{\mu\nu}=\epsilon\,\partial_\mu\phi\,\partial_\nu\phi-g_{\mu\nu}\left[\frac{\epsilon}{2}(\partial\phi)^2+V\right],\qquad
T^{(F)}_{\mu\nu}=\mathcal{L}_{\mathcal{F}}\,F_{\mu\alpha}F_\nu{}^{\alpha}-g_{\mu\nu}\mathcal{L},
\label{eq:Tmn}
\end{equation}
where $\mathcal{L}_{\mathcal{F}}=\dd\mathcal{L}/\dd\mathcal{F}$. Varying with respect to $\phi$ and $A_\mu$ gives the matter equations
\begin{equation}
\epsilon\,\Box\phi=\frac{\dd V}{\dd\phi},\qquad \nabla_\mu\!\left(\mathcal{L}_{\mathcal{F}}F^{\mu\nu}\right)=0,
\label{eq:matter}
\end{equation}
and the contracted Bianchi identity guarantees $\nabla^\mu T_{\mu\nu}=0$ once Eqs.~(\ref{eq:matter}) hold. We take the static ansatz $\phi=\phi(r)$ and the purely magnetic field $F_{\theta\phi}=q\sin\theta$, for which $\mathcal{F}=q^2/(2R^4)$ with $R^2=r^2+a^2$; the Maxwell-type equation in~(\ref{eq:matter}) is then satisfied identically, and the mixed components of the stress tensor are $T^t{}_t=-\tfrac{\epsilon}{2}A\phi'^2-V-\mathcal{L}$, $T^r{}_r=+\tfrac{\epsilon}{2}A\phi'^2-V-\mathcal{L}$ and $T^\theta{}_\theta=T^\phi{}_\phi=-\tfrac{\epsilon}{2}A\phi'^2-V-\mathcal{L}+2\mathcal{F}\mathcal{L}_{\mathcal{F}}$, with $\phi'=\dd\phi/\dd r$.

With the sectional curvatures of Sec.~\ref{isec2} the mixed Einstein tensor is $G^t{}_t=-(2E_3+E_4)$, $G^r{}_r=-(2E_2+E_4)$ and $G^\theta{}_\theta=-(E_1+E_2+E_3)$. The difference of the first two equations isolates the scalar field, since $E_2-E_3=AR''/R$,
\begin{equation}
\epsilon\,\phi'^2=-\frac{R''}{4\pi R}=-\frac{a^2}{4\pi R^4}\quad\Longrightarrow\quad
\epsilon=-1,\qquad \phi(r)=\frac{1}{\sqrt{4\pi}}\arctan\frac{r}{a},
\label{eq:phi}
\end{equation}
so the scalar must be phantom for any $a\neq0$, exactly as in the Simpson-Visser case, and its profile is independent of $\alpha$ and of $M$ because the bounce is entirely carried by $R(r)$. The remaining two combinations determine $\mathcal{L}$ and $V$. Subtracting the $tt$ from the $\theta\theta$ equation gives $2\mathcal{F}\mathcal{L}_{\mathcal{F}}=(E_3+E_4-E_1-E_2)/(8\pi)$, and since $\dd\mathcal{F}/\dd R=-4\mathcal{F}/R$ this is a first-order equation for $\mathcal{L}$ as a function of $R$ whose solution, with the integration constant fixed by $\mathcal{L}\to0$ at infinity, is
\begin{equation}
\mathcal{L}(R)=\frac{3Ma^2}{20\pi R^5}+\frac{3\alpha M^2\left(R^2-a^2\right)}{8\pi R^8},\qquad
\mathcal{L}(\mathcal{F})=\frac{3Ma^2}{20\pi}\left(\frac{2\mathcal{F}}{q^2}\right)^{5/4}
+\frac{3\alpha M^2}{8\pi}\left[\left(\frac{2\mathcal{F}}{q^2}\right)^{3/2}-a^2\left(\frac{2\mathcal{F}}{q^2}\right)^{2}\right].
\label{eq:LF}
\end{equation}
The sum of the $tt$ and $rr$ equations then fixes the potential, $V+\mathcal{L}=(E_2+E_3+E_4)/(8\pi)$, and using $R=a\sec(\sqrt{4\pi}\,\phi)$ from Eq.~(\ref{eq:phi}),
\begin{equation}
V(R)=\frac{Ma^2\left(4R^3-5\alpha M\right)}{40\pi R^8},\qquad
V(\phi)=\frac{M}{10\pi a^3}\cos^5\!\big(\sqrt{4\pi}\,\phi\big)-\frac{\alpha M^2}{8\pi a^6}\cos^8\!\big(\sqrt{4\pi}\,\phi\big).
\label{eq:Vphi}
\end{equation}
We verified by direct substitution that Eqs.~(\ref{eq:phi})-(\ref{eq:Vphi}) satisfy all three independent Einstein equations identically, and that the scalar equation in~(\ref{eq:matter}) follows from them by the Bianchi identity. At $\alpha=0$ the results reduce to the Bronnikov-Walia source of the Simpson-Visser spacetime, $\mathcal{L}\propto\mathcal{F}^{5/4}$ and $V\propto\cos^5$~\cite{Bronnikov:2022}. At $a=0$ the scalar switches off and the qOSD exterior is supported by the single power $\mathcal{L}=\tfrac{3\alpha M^2}{8\pi}(2\mathcal{F}/q^2)^{3/2}$, whose energy density $\rho=\mathcal{L}=3\alpha M^2/(8\pi R^6)$ is the one quoted in Sec.~\ref{isec2}; the holonomy correction is thus equivalent, at the level of the exterior geometry, to a magnetic nonlinear electrodynamics with a $\mathcal{F}^{3/2}$ Lagrangian. In the general case the two sectors mix only through the last term of Eq.~(\ref{eq:LF}) and the last term of Eq.~(\ref{eq:Vphi}), both proportional to $\alpha a^2$, and the magnetic charge $q$ is a free normalization that drops out of the metric. The negative sign of the phantom kinetic term is the field-theoretic counterpart of the null-energy violation of Eq.~(\ref{eq:nec0}): a canonical scalar cannot produce $R''>0$ at a throat, whatever the potential or the electrodynamics.

\bibliographystyle{unsrtnat_linkizzet}

\bibliography{finalref}

\end{document}